\documentclass[
 reprint,
 amsmath,amssymb,
 aps,
]{revtex4-2}

\usepackage{graphicx}
\usepackage{dcolumn}
\usepackage{bm}
\usepackage{xcolor}
\usepackage{tikz}

\definecolor{lime}{HTML}{A6CE39}
\DeclareRobustCommand{\orcidicon}{
	\begin{tikzpicture}
	\draw[lime, fill=lime] (0,0) 
	circle [radius=0.16] 
	node[white] {{\fontfamily{qag}\selectfont \tiny ID}};
	\draw[white, fill=white] (-0.0625,0.095) 
	circle [radius=0.007];
	\end{tikzpicture}
	\hspace{-2mm}
}
\foreach \x in {A, ..., Z}{\expandafter\xdef\csname orcid\x\endcsname{\noexpand\href{https://orcid.org/\csname orcidauthor\x\endcsname}{\noexpand\orcidicon}}}

\usepackage[font=small,labelfont=bf,justification=justified]{subcaption}
\usepackage[font=small,labelfont=bf,justification=justified]{caption}

\usepackage{hyperref}
\usepackage{ulem}

\begin{document}

\preprint{APS/123-QED}

\title{Influence of the crust on the neutron star macrophysical quantities and universal relations}
\author{L. Suleiman\orcidA{}}
\affiliation{Facult\'{e} des Sciences, Parc de Grandmont, 37200 Tours, France}
\affiliation{Nicolaus Copernicus Astronomical Center of the Polish Academy of Sciences, ul. Bartycka 18, 00-716 Warszawa, Poland}
\affiliation{LUTH, Observatoire de Paris, Universit\'{e} PSL, CNRS, Universit\'{e} de Paris, 92190 Meudon, France}

\author{M. Fortin\orcidB{}}
 \affiliation{Nicolaus Copernicus Astronomical Center of the Polish Academy of Sciences, ul. Bartycka 18, 00-716 Warszawa, Poland}

\author{J. L. Zdunik\orcidC{}}
 \affiliation{Nicolaus Copernicus Astronomical Center of the Polish Academy of Sciences, ul. Bartycka 18, 00-716 Warszawa, Poland}
\author{P. Haensel\orcidD{}}
\affiliation{Nicolaus Copernicus Astronomical Center of the Polish Academy of Sciences, ul. Bartycka 18, 00-716 Warszawa, Poland}

\date{\today}

\begin{abstract}
\begin{description}
\item[Background] 
Measurements of neutron-star macrophysical properties thanks to multimessenger observations offer the possibility to constrain the properties of nuclear matter. 
Indeed cold and dense matter as found inside neutron stars, in particular in their core, is not accessible to terrestrial laboratories.

\item[Purpose] 
We investigate the consequences of using equations of state that employ models for the core and the crust that are not calculated consistently on the neutron-star macrophysical properties, on some of the so-called universal relations and on the constraints obtained from gravitational-wave observations.

\item[Methods] 
We use various treatments found in the literature to connect together nonconsistent core and crust equations of state. We then compute the mass, the radius, the tidal deformability, and the moment of inertia for each model.
Finally, we assess the discrepancies in the neutron-star macrophysical properties obtained when consistent models for the whole star and nonconsistent ones are employed.

\item[Results]
The use of crust models nonconsistent with the core introduces an error on the macrophysical parameters which can be as large as the estimated accuracy of current and next generation telescopes. 
The precision of some of the universal relations reported in the literature is found to be overestimated. We confirm that the equation of the crust has limited influence on the macrophysical properties.

\item[Conclusions]
The discrepancy between results obtained for a fully consistent equation of state and a nonconsistent one can be reduced if one connects the core and the crust models at baryon densities around 0.08-0.1 fm$^{-3}$.
The equation of the crust cannot be probed with current multimessenger observations and near-future ones.

\end{description}
\end{abstract}

\maketitle

\section{Introduction}

Very few unified equations of state (EOS) in the sense that the same nuclear model is used to describe the whole interior of the neutron star (NS), typically the core and the crust, are available. 
This mostly originates from the fact that, while the core is homogeneous, the crust has a crystalline structure whose modeling is not as straightforward. Various techniques are used to connect the core and crust EOS, however very often the resulting EOS is not thermodynamically consistent.
Hence artificial uncertainties in the NS radius $R$ arise and those can be as large as the precision expected from current and future X-ray telescopes: NICER \citep{NICER,Watts16}, Athena \citep{Athena}, and eXTP \citep{eXTP} as discussed in, e.g. \cite{Fortin16}. 
Using a unified EOS prevents this type of thermodynamical inconsistencies and hence prevents the introduction of uncertainties when calculating the NS macrophysical parameters, for example the radius. 
Alternatively, some approximate approaches to the crust, such as in Ref.~\cite{Zdunik17} allow for calculation of the radius of a NS with a very good precision while employing only the core EOS. 
Recently, the first simultaneous determination of the radius and mass $M$ of a NS with the NICER mission was obtained after modeling the pulsating X-ray emission from the isolated millisecond pulsar PSR J0030$+$0451 \citep{NICERa,NICERc}. 
However, due to several difficulties in modeling the sources, the uncertainties are still quite large: ${R}={13.02}_{-1.06}^{+1.24}$ km and $M={1.44}_{-0.14}^{+0.15}\,M_{\odot}$  (68\% uncertainty) \citep{NICERc}, and $M={1.34}_{-0.16}^{+0.15}\,M_{\odot}$ and  $R={12.71}_{-1.19}^{+1.14}$ km \citep{NICERa}.

\begin{table*}
\caption{Nuclear properties of the various core and crust EOS. Energy per nucleon ($E_{\rm s}$), compression modulus ($K$), symmetry energy ($J$), and slope of the symmetry energy ($L$) at saturation density $n_{\rm s}$ for uniform symmetric nuclear matter. The density at the interface between the core and the crust is denoted as $n_{\rm t}$ \citep{Ducoin11,Pearson12,Fortin16}. The last two columns give the value of the symmetry energy and its slope at the density 0.1 fm$^{-3}$:  $J_{0.1}$ and $L_{0.1}$, respectively.}
\label{tab:eos}
\begin{tabular}{lcccccc|cc}
\hline
\hline
Model& $n_{\rm s}$       & $E_{\rm s}$ & $K$  & $J$  & $L$ &$n_{\rm t}$& $J_{0.1}$  & $L_{0.1}$\\
       & (fm$^{-3}$) &(MeV) & (MeV)& (MeV)&(MeV) &(fm$^{-3}$) & (MeV)&(MeV)  \\
\hline

Core & & & & &  & & &\\
NL3                 & 0.149 & -16.2 & 271.6 & 37.4 & 118.9 & 0.057 & 25.0 & 73.7\\
BSR6                & 0.149 & -16.1 & 235.8 & 35.6 &  85.7 & 0.061 & 25.8 & 62.9\\
DD2                 & 0.149 & -16.0 & 242.6 & 31.7 &  55.0 & 0.067 & 24.9 & 70.1\\
\hline
Crust & & & & &  & & &\\
DH                  & 0.159 & -16.0 & 230.0 & 32.0 & 46.0  & 0.076 & 25.2 & 41.6\\
BSk21               & 0.159 & -16.1 & 245.8 & 30.0 & 46.6  & 0.081 & 23.7   & 36.8  \\
\hline
\hline
\end{tabular}
\end{table*}

As far as the uncertainty on the moment of inertia $I$ and the tidal deformability $\Lambda$ are concerned, the influence of the matching (or gluing) between the core and crust EOS has hardly been studied. However, the tidal deformability of the NS composing the binary system that emitted the gravitational wave (GW) signal GW170817 has been constrained and more measurements with better precision are expected during new observational runs of the LIGO-Virgo-KAGRA collaboration. 
No moment of inertia $I$ of NS has been measured so far.
Double pulsars are particularly suitable for such measurements due to the precision of the observations and the extreme nature of the system. So far only one of such system is known:  PSR J0737$-$3039 \citep{Lyne04}, with the two pulsars visible until pulsar B radio disappeared in 2008 due to precession. 
However with more observations with current instruments and the FAST \citep{FAST} and SKA \citep{SKA} radiotelescopes, the number of known pulsars is expected to increase by orders of magnitude, including thousands of millisecond pulsars, and among them possibly binary systems with two pulsars. 
One may then be able to determine the moment of inertia of some NS. 

This paper assesses the influence of the matching between an EOS for the core and another for the crust on the determination of the radius, tidal deformability and moment of inertia, which may help us to better constrain the properties of the matter inside NSs when these quantities  will be measured precisely.  We also check to which extent various so-called universal relations, fits obtained between $I$, $\Lambda$ and the NS compactness $C$ are affected by the core-crust matching. Finally we study the influence of the crust EOS itself on the NS macrophysical quantities and assess whether it is possible to gain insight on the NS crust from multi-messenger observations.

\begin{figure*}
\resizebox{0.68\hsize}{!}{\includegraphics{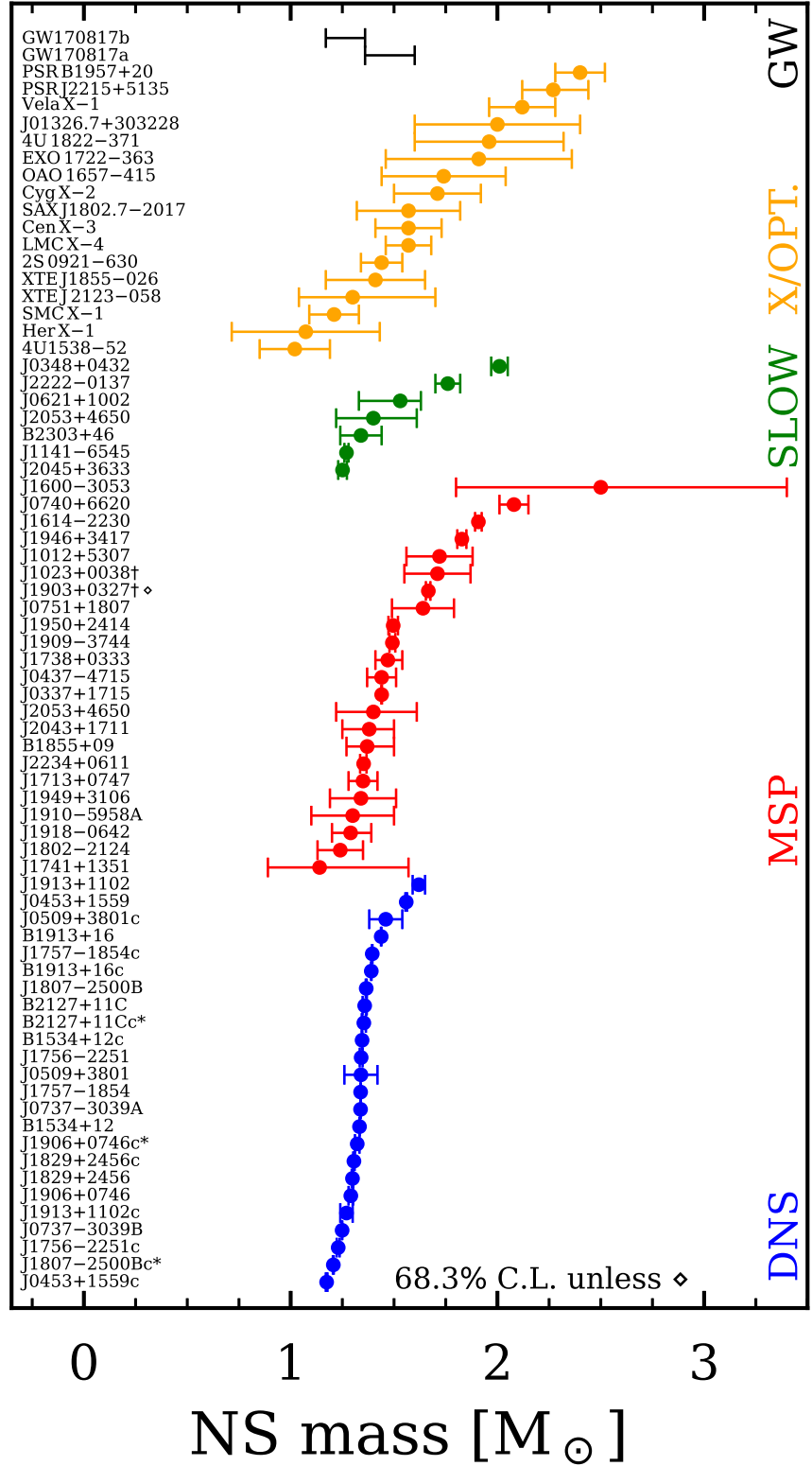}}
\caption{Mass measurements of 73 NS within $68.3\%$ confidence level (1$\sigma$) except for PSR J1903+0327: 99.7\% ($\diamond$ symbol) and GW170817. The following classification is adopted: binaries with two neutron stars (DNS), millisecond pulsars (MSP) with spin frequency $f\geq 50$Hz and with a companion that is not a NS, slowly rotating pulsars (SLOW) with spin frequency $f\leq50$Hz not in a DNS, X/OPT for NSs measured through X-ray or optical observations (as opposed to radio in previous categories) and GW for NS mass measurements using detection of gravitational waves. Data from \cite{Freire} (Jan. 2021), \cite{Ozel2016}, Table 1 in \cite{Alsing2018}, \cite{Lattimer2012} and \cite{GW170817}.}
\label{Fig:masses}
\end{figure*}

\section{Equation of state construction}

We start by employing three EOS for a purely nucleonic core (uniform $npe\mu$ mixture) obtained from RMF calculations for the NL3, BSR6, and DD2 parametrizations \citep{Fortin16}. 
Table \ref{tab:eos} gives for these models various nuclear properties at saturation and at 0.1~fm$^{-3}$ together with the value of the density $n_{\rm t}$ at the transition between the core and the crust. 
NL3 is the stiffest model while DD2 is the softest one in the sense that at a given density NL3 gives the largest pressure and DD2 the lowest. 
BSR6 stands in the middle with a moderate stiffness. 

So far the only robust astrophysical constraint on NS matter comes from mass measurements. 
Indeed the maximum mass obtained for a given EOS has to be larger than the largest observed mass - otherwise the EOS is not consistent with observations and can be ruled out.
In Fig.~\ref{Fig:masses}, we present 73 NS mass measurements divided into four categories. 
The most accurate determinations of pulsar masses are based on the measurement of at least two post-Keplerian (PK) parameters (in addition to the classical Keplerian ones) in Double Neutron Stars (DNS) systems. 
In such binaries only one of the two components is observed as a pulsar (with the exception of J$0737-3039$ where the two NSs were observed as pulsars until 2008).
Whether it is possible or not to measure a specific PK parameter in a binary system depends on the shape, size and orientation of the orbit. 
For example, the rate of periastron advance is measurable for eccentric orbits, and Shapiro delay parameters for rather large companion mass with an edge-on orientation of the orbit. 
Most of the millisecond pulsars (MSPs) in binaries have a white dwarf as a companion. 
The determination of the pulsar mass is, in these cases, mainly based on Shapiro delay measurements, but there exist systems in which the spectroscopic observations of white dwarfs provides a classical measurement of the orbit parameters needed to determine the mass of each star. 
Also, in X-ray binary systems the analysis of the optical observations of the companion is crucial to estimate the NS mass. 
The uncertainty in these cases is significantly larger than in the case of all DNS and many MSP binaries and very often the systematic uncertainty dominates. 
So far the largest observed mass with a good precision is the one of J0740$+$6620: $M=2.08\pm 0.07\,M_{\odot}$, \cite{fonseca2021refined}. It should be noted, however, that analysis of kilonova GW170817 suggests an upper limit on the maximum mass of NSs at $2.17\,M_\odot$ \cite{margalit2017}, although the value larger by $ \approx 0.2\,M_\odot$ is also possible \cite{shibata2019}.
The three EOSs for the core considered in this work are all consistent with this mass constraint because they have the respective maximum masses: $M^{\rm NL3}_{\rm max} =2.77\,M_{\odot}$, $M^{\rm DD2}_{\rm max} = 2.42\, M_{\odot}$ and $M^{\rm BSR6}_{\rm max} = 2.43\,M_{\odot}$.

Regarding nuclear properties, note that only DD2 is consistent with experimental constraints on parameters of neutron-star matter, in particular the symmetry energy and its slope at saturation (see discussion in Ref.~\cite{Fortin16}). 
We nevertheless include the stiff BSR6 and the very stiff NL3 models in order to study to which extent the use of nonunified EOSs creates artificial uncertainties in theoretically calculated $R$, $I$, and $\Lambda$. 

For the low-density part of the EOS, we use:
\begin{itemize}
\item a crust calculated consistently with the core, that is using the same nuclear parametrization, from Ref.~\cite{Fortin16}. The resulting EOS is then a quasi-unified EOS because the outer crust is not calculated consistently with the inner crust and core. However it was shown in Ref.~\cite{Fortin16} that the radius is hardly affected when a nonconsistent outer crust is used in the sense that the uncertainty that is introduced is much less than the precision of any current or near-future measurements,
\item the DH EOS based on the SLy4 Skyrme force parametrization from Ref.~\cite{Douchin01},
\item the BSk21 model for a catalyzed (that is nonaccreted) crust from Ref.~\cite{Fantina13},
\item the BSk21 model for a fully-accreted crust recently obtained in Ref.~\cite{Fantina18}.
\end{itemize}

Their nuclear properties are also shown in Table \ref{tab:eos}. 
The crust of isolated NSs is described by models for a catalyzed crust. 
However, when they are in binary systems, NS may undergo some periods of accretion of the matter from their companion star. 
The accreted matter falls onto the star at the surface of the crust and will then be pushed inside the crust as more matter is accreted. 
As the accreted matter sinks in the star, the crust will undergo a series of nuclear reactions changing its composition and thus become a so-called accreted crust.
During the evolution of a binary neutron-star system, it may be that one of the two NSs accretes matter from the other star before the latter ends its life in a supernova explosion \cite{Tauris_2017}. 
Therefore, it is interesting to consider the influence of the nature of the crust, accreted or catalyzed, in particular in the context of detections of GWs from binary NS mergers.

The core-crust transition at the density $n_{\rm t}$ corresponds to the point where uniform matter becomes  unstable with respect to the spatial variations in the particle densities. A linear dependence of the transition density on the slope of the symmetry energy at saturation density was proposed in Refs.~\cite{tsang2019, Pais_2016}. Various techniques exist that can be used to determine $n_{\rm t}$ in NSs, but uncertainty in its value is large. For example, for the broad collection of EOS considered in Refs.~\cite{Oyamatsu_2007,Ducoin11,Pais_2016}, it has been  found  that $n_{\rm t}$  ranges between $0.05$ and $0.09$~fm$^{-3}$ that is $(0.3-0.6)n_0$ with the nuclear saturation density $n_0=0.16$~fm$^{-3}$.
For many dense-matter models used to describe NS interiors, $n_{\rm t}$ was not calculated at all, hence a widely used approach when constructing an EOS for the whole NS, is to connect (glue) the core EOS to the crust EOS. 
Various crust-core matching procedures are used, see discussion in \cite{Fortin16} and we study their influence on the determination of the NS macrophysical properties.

\section{Influence of the core-crust transition density}

\begin{figure*}
\begin{subfigure}[b]{0.49\hsize}
\resizebox{\hsize}{!}{\includegraphics{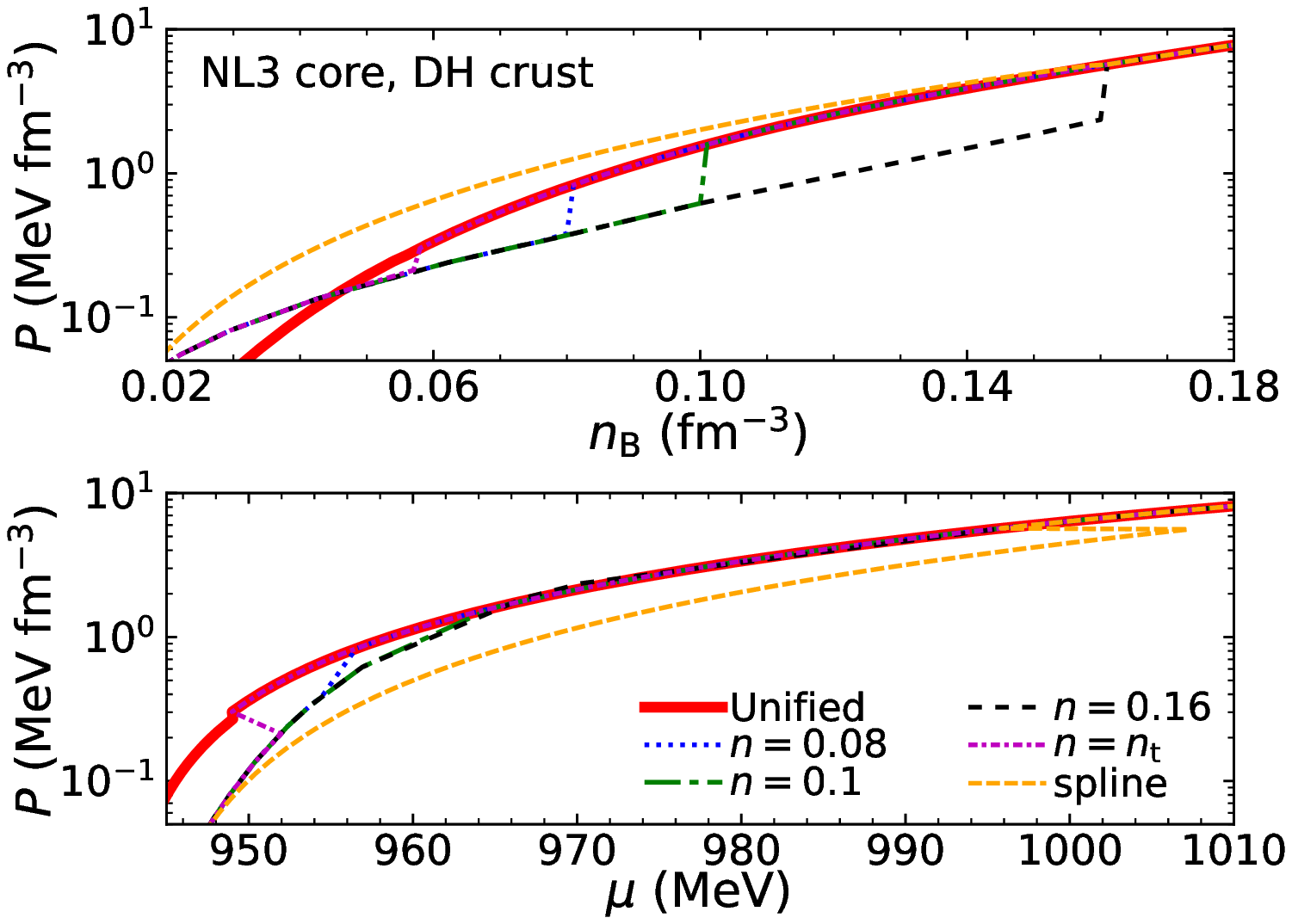}}
\caption{Matching of the pressure $P$ between the DH crust and the NL3 core as a function of the baryon number density $n_{\rm B}$ (upper plot) and the chemical potential $\mu$ (lower plot). }
\label{Fig:EOS_NL3}
\end{subfigure}
\hfill
\begin{subfigure}[b]{0.50\hsize}
\resizebox{\hsize}{!}{\includegraphics{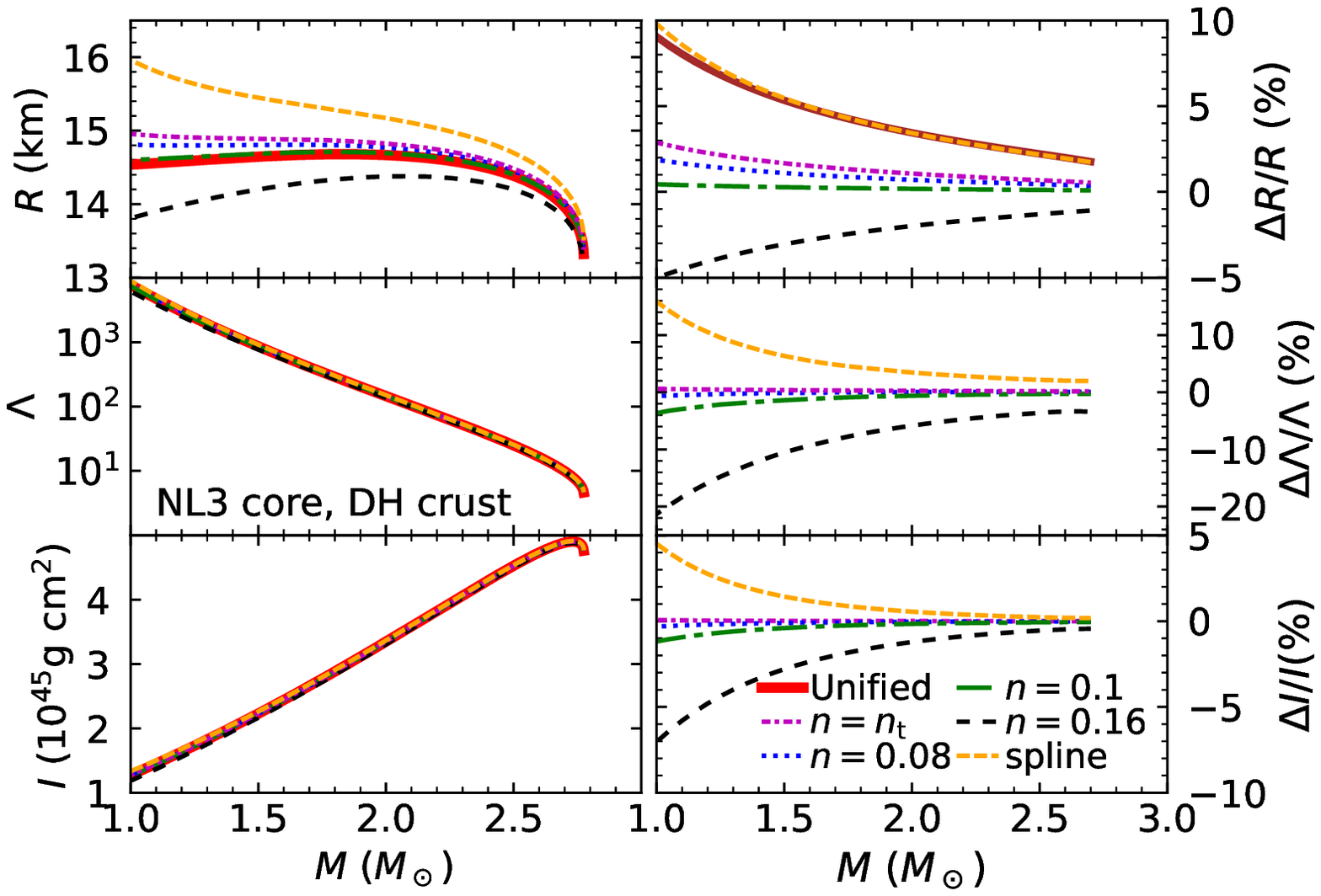}}
\caption{$M$, $R$, $I$ and $\Lambda$ for the matched and unified EOS  (left panels) and the relative differences with respect to the unified EOS for the NL3 core  (right panels). The brown line on top of the one for the spline shows the results obtained using the approximate approach to the crust (see text for details).}
\label{Fig:Delta_NL3}
\end{subfigure}
\caption{Matchings between the crust and the NL3 core EOS and ensuing uncertainties on the macrophysical parameters. The red thick line corresponds to the unified NL3 EOS and the others to different gluings of the DH crust EOS to the NL3 core EOS. }
\end{figure*}

In this section, the radius, moment of inertia and tidal deformability obtained using a unified EOS are compared with those calculated when the DH EOS is "glued" to the core EOS at various baryon-number densities $n_{\rm B}$: at the core-crust transition $n_{\rm t}$ calculated for the RMF models of the core (see Table~\ref{tab:eos}), at $n_0$, $n_0/2$, 0.10 fm$^{-3}$.
We also consider a matching employed in, e.g., \cite{Margueron18} that uses a cubic spline of the logarithm of the pressure $P$ in terms of the logarithm of the mass-energy density $\rho$  over the range of densities between 0.1$n_0$ and $n_0$. 
These "matched" EOS form a representative set that covers a large number of constructions used in the literature and in the NS community to calculate NS properties for a given core EOS. For the matching with a spline, we recalculate the density $n_{\rm B}$ using the relation:
\begin{equation}
    \frac{{\rm d} n_{\rm B}}{n_{\rm B}}=\frac{{\rm d}\rho}{P/ c^2+\rho}
\end{equation}
with  $c$ being the speed of light. 
For the other matched EOS, we use $n_{\rm B}$ as given by either the core or the crust EOS. 
Finally, for all EOS, matched or not, we compute the chemical potential $\mu=(P+\rho c^2)/n_{\rm B}$.

\subsection{NL3}

In principle, when gluing two EOSs one should match all thermodynamical quantities: $P$, $\rho$, and $n_{\rm B}$.
In other words, a pair of functions: $P(n_{\rm B})$ and $\rho(n_{\rm B})$ should be constructed so that thermodynamical consistency is fulfilled, ensuring that the chemical potential as a function of the pressure is continuous.

Figure.~\ref{Fig:EOS_NL3} shows the pressure $P$ as a function of the baryon number density $n_{\rm B}$ and as a function of the chemical potential $\mu$ for the different matchings between the core and the crust EOS that we consider. 
We observe a jump in the pressure as a function of the baryon number density at the transition between the core and the crust for all matched EOSs except the one employing a spline. 
All matched EOS exhibit a jump in the chemical potential. In the case of the gluing with a spline, the jump is due to the fact that even if the pressure is continuous, the mass-energy density is not at the upper bound of the interpolated crust EOS, at $n_{\rm B}=n_0$. 

Once we constructed an EOS for the whole NS (crust and core) we calculate macrophysical properties; in this work we only focus on nonrotating NSs.
We solve the Tolman-Oppenheimer-Volkoff equations for hydrostatic equilibrium in General Relativity to obtain the NS radius as a function of star mass. 
For a given mass we calculate the moment of inertia following the approach presented in \cite{HPY} and the tidal deformability as in, e.g., \cite{Malik18}.

The fact that the core and crust EOS are not glued together in a thermodynamically consistent way results in different values of the radius, moment of inertia and tidal deformability as a function of the mass compared with the unified EOS.
The  left panels of Fig.~\ref{Fig:Delta_NL3} show $R$, $\Lambda$, and $I$ as a function of $M$ for the different matchings between the NL3 core and the crust and for the unified NL3 EOS. 
We restrict ourselves to masses larger than $1.0\,M_\odot$ as the current measured masses range from $ \approx 1.2$ to more than $2\,M_\odot$, see Fig.~\ref{Fig:masses}. 
One can clearly see that there are differences, sometimes large, in the $M(R)$ relation depending on the matching. 
To quantify the effect of the matching, we follow the approach already used in \cite{Fortin16} and calculate for a given $M$ and variable $X=R,\Lambda,I$ the relative difference between the variable for a given matching $X_{\rm m}$ and for the unified EOS $X_{\rm u}$: $\Delta X/X=(X_{\rm m}-X_{\rm u})/X_{\rm u}$. 
The results are plotted in the right panels of Fig.~\ref{Fig:Delta_NL3}.

The matching between the core and  the crust introduces a relative difference with respect to the unified EOS in the radius determination as large as $ \approx 5\%$ in absolute value. 
In the case of the spline presented in Fig.~\ref{Fig:Delta_NL3} it is even as large as 10\%, or 1.5~ km for $M \approx1\,M_\odot$. 
The inaccuracy due to this discontinuity in $\mu$ can be estimated by the formula from \cite{Zdunik17}:
\begin{equation}
    \Delta R/R= - 0.72\%\, \frac{\Delta\mu}{\rm 1\,MeV} \frac{R}{\rm 10\, km} \frac{M_{\odot}}{M}(1-2GM/Rc^2)
\end{equation}
where $\Delta\mu$ is the difference between the chemical potential at the core-crust transition and at the surface, and $R$ is the radius at given mass $M$,  obtained for the matched EOS. 
The validity of this approximation for the "spline" matching is presented by the brown solid line in the left panel of Fig.~\ref{Fig:Delta_NL3}. 
The sign of the relative difference is related to the sign of the jump in the chemical potential as can be seen in the bottom panel in Fig.~\ref{Fig:EOS_NL3}: a drop in $\mu$ results in a radius being larger and hence a positive $\Delta R/R$, and vice versa. 
The inaccuracy in $R$ due to the discontinuity in $\mu$ corresponds to a similar formula for the compactness $C=GM/Rc^2$: 
\begin{equation}
    \Delta C =  \left(\Delta\mu/\mu_0\right)(1-2C)
\end{equation}
with $\mu_0$ being the chemical potential at the surface of the star (where $n_{\rm B}=0$). 
The relative error $\Delta C/C$ is larger than  $ \approx 5\%$ in the most extreme case of large $\Delta\mu \approx 10\,{\rm MeV}$ and relatively small $C \approx 0.1$.
Similarly, the thermodynamical inconsistency in the matched EOS results in relative differences with respect to the unified EOS in the tidal deformability and moment of inertia with values up to $ \approx20\%$ and $ \approx10\%$, respectively. 
We also observed that, as the mass increases, the discrepancy in the macrophysical quantities between the glued EOS and the unified one decreases. 
This is because the crust contribution to the NS macrophysical properties $R$, $\Lambda$, and $I$ becomes smaller relative to that of the core as $M$ increases. 
The uncertainties are the smallest when the core and crust are connected at $n_0/2=0.08$ fm$^{-3}$ or $0.1$~ fm$^{-3}$. 
We expect that the uncertainties due to the core-crust gluing are the largest for the NL3 core EOS as it has values of the symmetry energy and slope that are the most different from those of the DH crust model as seen in Table~\ref{tab:eos}. 
NL3 is the stiffest EOS of all models while the crust EOS employed in this section, the DH one, is much softer. 
\begin{figure*}
\begin{subfigure}[b]{0.49\hsize}
\resizebox{\hsize}{!}{\includegraphics{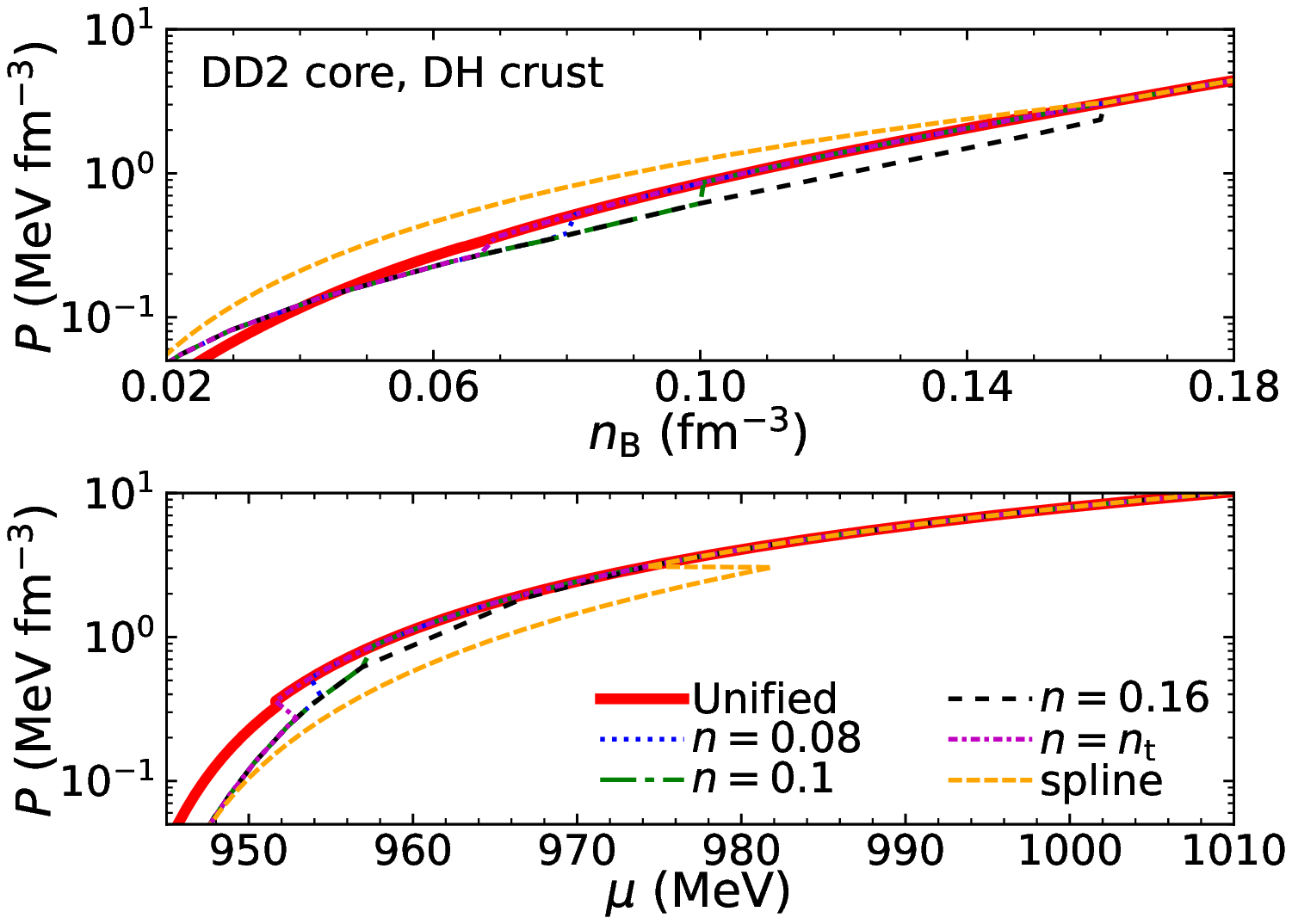}}
\caption{Pressure $P$ as a function of the baryon number density $n_{\rm B}$ (upper plot) and the chemical potential $\mu$ (lower plot) for the various matched and unified EOS with the DD2 core.}
\label{Fig:EOS_DD2}
\end{subfigure}
\hfill
\begin{subfigure}[b]{0.50\hsize}
\resizebox{\hsize}{!}{\includegraphics{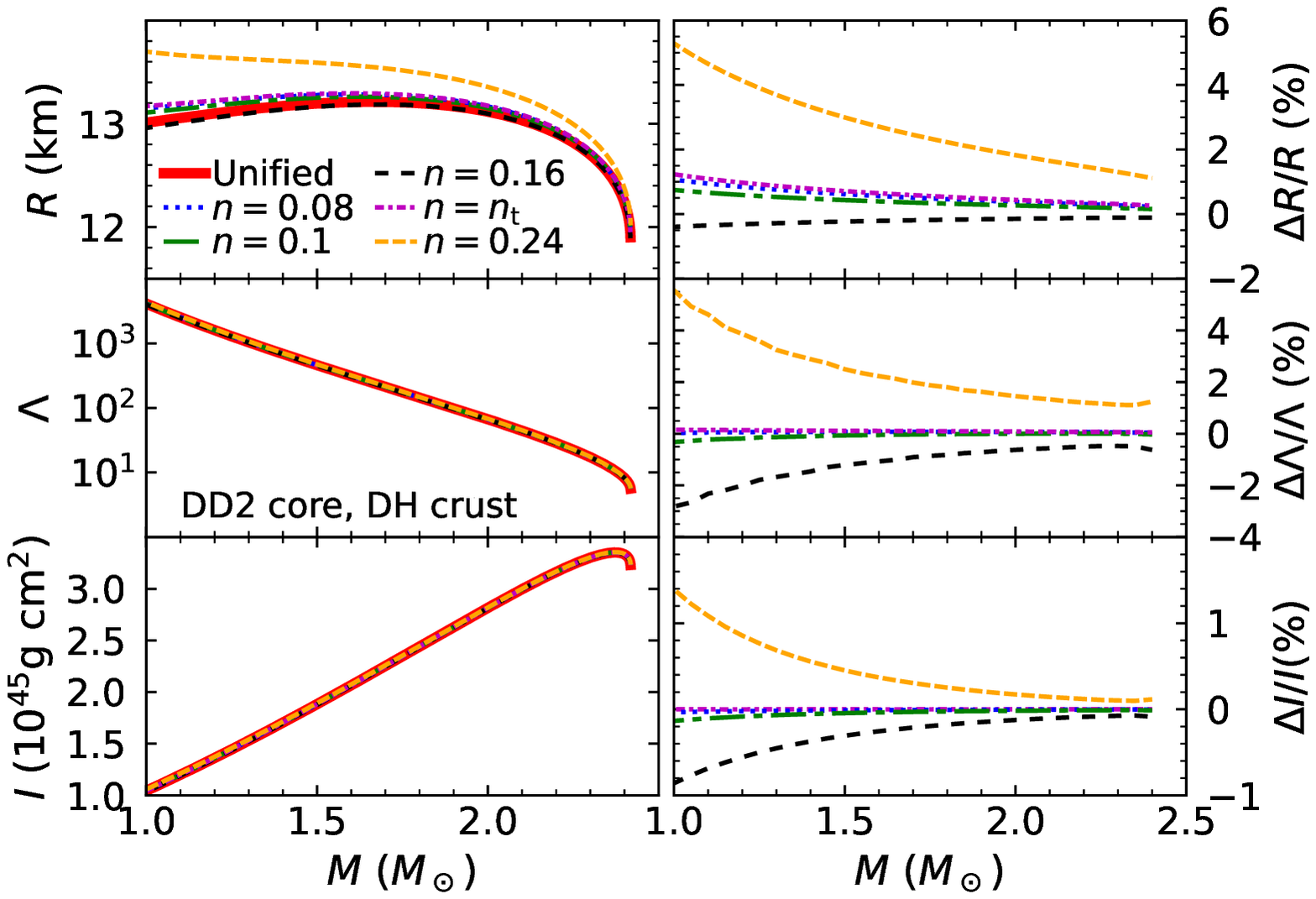}}
\caption{$M$, $R$, $I$ and $\Lambda$ for the various matched and unified EOS (left) and the relative differences with respect to the unified EOS DD2 (right).}
\label{Fig:Delta_DD2}
\end{subfigure}
\caption{Matchings between the crust and the DD2 core EOS  and ensuing uncertainties on macrophysical parameters.}
\end{figure*}

\subsection{BSR6 and DD2}

We use a similar approach for the BSR6 and DD2 core EOS. The constructions of DD2 EOS are shown in Fig.~\ref{Fig:EOS_DD2} and in the Appendix in Fig.~\ref{Fig:EOS_BSR6} for BRS6 EOS. 
The macrophysical properties and uncertainties due to the matching are shown in Fig.~\ref{Fig:Delta_DD2} and in the Appendix in Fig.~\ref{Fig:Delta_BSR6} for DD2 and BSR6, respectively. 
As a consequence of the thermodynamical inconsistency relative differences with respect to the unified EOS as large as (in absolute value) $ \approx$ $6\%$, $10\%$, and $3\%$ for the radius, tidal deformability and moment of inertia respectively are obtained for the BSR6 core and $5\%$, $6\%$, and $1.5\%$ for the DD2 EOS. 
The differences are smaller for the DD2 core model than for the BSR6 one because the former has a symmetry energy and its slope at saturation that are most similar to that of DH (see Table~\ref{tab:eos}). 

We note again that the discrepancies with respect to the unified EOS are minimized when the matching is performed at $0.08$ or $0.1$~fm$^{-3}$ or $ \approx 0.5n_0$, which is the value of the core-crust transition density found for a number of nuclear models \cite{Oyamatsu_2007,Ducoin11,Pais_2016}. In fact, laboratory experiments allow us to constrain relatively well the NS EOS up to roughly $n_0/2$, see e.g., Refs.~ \cite{Dutra_2012, Dutra_2013,Brown_2013}. Since most EOSs are adjusted to reproduce the experimental data, they consequently have properties that are similar up to $ \approx n_0/2$, for example the symmetry energy as can be seen in Figure 2 of Ref.~\cite{Ducoin11}.
From Table \ref{tab:eos} we can see that the models we consider have remarkably close values of the symmetry energy at a density $0.1$~fm$^{-3}$, around $25$ MeV. 
The spread of the slope of the symmetry energy is also lesser at $0.1$ fm$^{-3}$ than at the saturation density. 
In other words all the EOS that we use have similar softness around $0.1$~fm$^{-3}$ which is why the jump in the chemical potential when gluing them is small in the range of densities $0.08-0.1$~fm$^{-3}$. In the end, this results in relative differences for the macrophysical properties that are small as can be seen from Fig.~\ref{Fig:Delta_DD2} and Fig.~\ref{Fig:Delta_BSR6} of the Appendix.

\begin{table*}
\caption{Maximum relative difference in \% between the values obtained using the various fits discussed in this paper and the exact calculations for the three core EOS. We consider unified EOS (column "uni.") and three EOSs matched to the crust at different densities:  $n_0$, $n_0/2$ and 0.1~fm$^{-3}$. Numbers in parentheses correspond to the NS mass at which the relative difference is the largest.}
\label{tab:fit}
\centering
\begin{tabular}{c|cccc|cccc}
\hline 
\hline 
EOS & uni. & $n_0$ & $n_0/2$ & 0.1 & uni. & $n_0$ & $n_0/2$ & 0.1 \\
\hline 
 & \multicolumn{4}{c|}{} & \multicolumn{4}{c}{}\\
$\Lambda-C$ fits & \multicolumn{4}{c|}{Maselli et al.} & \multicolumn{4}{c}{Yagi and Yunes}\\
\hline 
NL3  & 3.65 (1.02)& 3.05 (1.01)& 5.53 (1.01)& 4.74 (1.00)& 2.94 (1.01)& 2.74 (1.01)& 1.85 (1.44)& 1.78 (1.00)\\
BSR6 & 5.57 (1.00)& 4.35 (1.00)& \textbf{6.38 (1.01)}& 5.77 (1.01)& 2.12 (1.33)& 0.92 (1.35)& \textbf{2.71 (1.30)}& 2.20 (1.31)\\
DD2  & 4.45 (1.00) & 4.54 (1.01)& 5.46 (1.00)& 5.14 (1.01)& 1.11 (2.22)& 1.17 (2.21)& 1.72 (1.23)& 1.52 (1.23)\\

\hline 
 & \multicolumn{4}{c|}{} & \multicolumn{4}{c}{}\\

 $\bar I-\Lambda$ fits& \multicolumn{4}{c|}{Maselli et al.} & \multicolumn{4}{c}{Yagi and Yunes}\\
\hline 
 NL3  & 7.18 (1.01)& 5.13 (1.01)&\textbf{ 7.30 (1.01)}& 7.23 (1.00)& 0.28 (2.67)& \textbf{0.38 (2.61)}& 0.26 (2.66)& 0.27 (2.67)\\
 BSR6 & 4.49 (2.26)& 4.56 (2.26)& 4.47 (2.26)& 4.49 (2.26)& 0.17 (2.32)& 0.24 (2.29)& 0.16 (2.33)& 0.17 (2.32)\\
 DD2  & 4.55 (2.24)& 4.55 (2.23)& 4.53 (2.24)& 4.53 (2.24)& 0.23 (2.30)& 0.23 (2.30)& 0.21 (2.30)& 0.22 (2.30)\\

\hline 
 & \multicolumn{4}{c|}{} & \multicolumn{4}{c}{}\\

 $\tilde{I}-C$ fits& \multicolumn{4}{c|}{Breu and Rezzolla} & \multicolumn{4}{c}{Zhao and Lattimer}\\
\hline 
NL3 & 4.44 (2.60)& 5.90 (2.37)& 5.33 (1.01)& 4.23 (2.63)& 2.90 (2.51)& 3.16 (2.40)& 4.64 (1.01)& 3.23 (1.00)\\
BSR6 & 4.60 (1.00)& 3.66 (2.14)& \textbf{6.00 (1.01)}& 4.85 (1.01)& 4.58 (1.00)& 2.36 (1.00)& \textbf{6.01(1.01)}& 4.93 (1.01)\\
DD2 &  4.01 (2.14) & 4.11 (2.12)& 4.04 (1.00)& 3.64 (2.19)& 2.69 (1.00)& 2.82 (1.01)& 4.44(1.00)& 3.91 (1.01)\\

\hline 
 & \multicolumn{4}{c|}{} & \multicolumn{4}{c}{}\\

 $\bar{I}-C$ fits& \multicolumn{4}{c|}{Breu and Rezzolla} & \multicolumn{4}{c}{Yagi and Yunes}\\
\hline 
NL3 &  3.98 (2.77)& \textbf{5.06 (2.77)}& 3.52 (2.77)& 3.83 (2.77)& 3.04 (2.51)& 3.03 (2.11)& 3.38(2.51)& 3.14 (2.51)\\
BSR6 & 1.66 (1.00)& 2.35 (2.40)& 3.14 (1.01)& 2.04 (1.01)& 3.15 (1.17)& 1.00 (1.14)& \textbf{4.35(1.14)}& 3.40 (1.14)\\
DD2 &  2.81 (2.40) & 2.89 (2.40)& 2.46 (2.41)& 2.56 (2.41)& 2.06 (2.32)& 2.00 (2.32)& 2.92(1.06)& 2.55 (1.06)\\
\hline
\hline
\end{tabular} 
\end{table*}
All in all, when constructing a NS EOS, in case a unified EOS is not available, gluing the core to the crust at $n_{\rm B}=0.08-0.1$~fm$^{-3}$ minimizes the relative differences with respect to the unified EOS and thus the artificial uncertainties in the radius, tidal deformability and moment of inertia.

\subsection{Effect of matching on the quality of {\bf universal relations}}

In the following, for clarity and simplicity we consider only the matchings of the crust to the core EOS at densities $0.5n_0=0.08$~fm$^{-3}$ and $0.1$~fm$^{-3}$ because it results in the smallest uncertainties of all the matchings we consider; we keep the matching at $n_0$ as a reference. Also we will only highlight plots for the stiffest and softest core EOS: NL3 and DD2, respectively. Indeed, the results for BSR6, which is softer than NL3 but stiffer than DD2, lie between those for the two extreme core EOS ; results for BRS6 EOS can be found in the Appendix. 

Several so-called universal relations between various NS macrophysical properties, e.g., $C$, $I$, $\Lambda$,  the f-mode oscillation frequency, have been reported in the literature - see \cite{Yagi17} for a review. They are called universal because they depend little on the EOS and their simple forms allow us to perform GW data analysis that would not be possible without them. However, their dependence on the EOS is of the order of a few percents. With current GW observatories, the systematic errors on the estimation of parameters using the universal relations are smaller than the statistical uncertainties. Nevertheless, for future and more precise detectors, more accurate relations will be needed \cite{Baiotti_2019}.

In the following we focus on relations between $C$, $\Lambda$, and $I$. Relations involving the quadrupole moment $Q$ are beyond the scope of this work.

\subsubsection{Relations between $\Lambda$ and $C$}

We focus on two universal relations between the tidal deformability and the compactness expressed as:
\begin{equation}
    C_{\rm fit}=\sum_{k=0}^{k=2}a_k(\ln\Lambda)^k
\end{equation}

The Maselli et al. fit  reported in \cite{Maselli13}  for masses in the range $1.2-2.0\,M_\odot$
has been obtained using only three nonunified purely-nucleonic EOS with a maximum mass larger than $2\,M_\odot$ and modeled by the piece-wise polytropic fits obtained in \cite{Read09}.  It yields $a_0=0.371$, $a_1=0.0391$, and $a_2=0.001056$ and gives a reported relative error $|C-C_{\rm fit}|/C_{\rm fit}$ of $\lesssim 2\%$.
 
The second fit, the Yagi \& Yunes one from \cite{Yagi17}, with $a_0=0.360$, $a_1=-0.0355$, and $a_2=0.000705$ has been obtained with a much larger set of $ \approx 30$ EOS for NS and quark stars (again piece-wise polytropic fits for NS)  and gives a reported maximum  error of 6.5\% for NS.

In Fig.~\ref{Fig:ICLove} in the Appendix, the top panels show results for the softest of our core models: DD2 (right side) and the stiffest one: NL3 (left side), the relations between $\Lambda$ and $M$. For each nonunified and unified EOS we calculate $C$ and compare it to the value $C_{\rm fit}$ obtained using  the relations derived in Refs.~\cite{Maselli13,Yagi17}. In the middle panels of Fig.~\ref{Fig:ICLove}  we show the relative error  $|C-C_{\rm fit}|/C_{\rm fit}$ with respect to $\Lambda$. For each matching,  EOS and  fit, the value of the largest relative error is indicated in Table \ref{tab:fit} together with the mass at which this difference is obtained. We do not show plots for the BSR6 core but results are included in the discussion.

Overall the Yagi \& Yunes fit gives  a smaller relative error ($ \approx 3\%$ at most) than the Maselli et al. fit (up to $ \approx 6\%$). The maximum difference is larger for the Maselli et al. fit than for the Yagi \& Yunes one for stars with masses $\lesssim 2\,M_\odot$ while the situation is opposite for larger masses. Table \ref{tab:fit} shows that the use of a nonconsistent EOS matched at $n_0=0.08$ fm$^{-3}$ or $0.1$ fm$^{-3}$ gives rise to an uncertainty which is smaller than the reported precision of the Yagi and Yunes fit but larger by up to a factor $ \approx 3$ for the Maselli et al. fit. Actually, for the latter fit, the relative difference in the compactness  when using a unified EOS is about two times larger than the reported precision. Hence using three EOSs covering a large range of stiffness we conclude that results obtained with the Yagi \&  Yunes fit are not affected by the treatment of the low-density part of the EOS and the matching while the Maselli et al. fit is.

\begin{figure*}
\begin{subfigure}[b]{0.47\hsize}
\resizebox{\hsize}{!}{\includegraphics{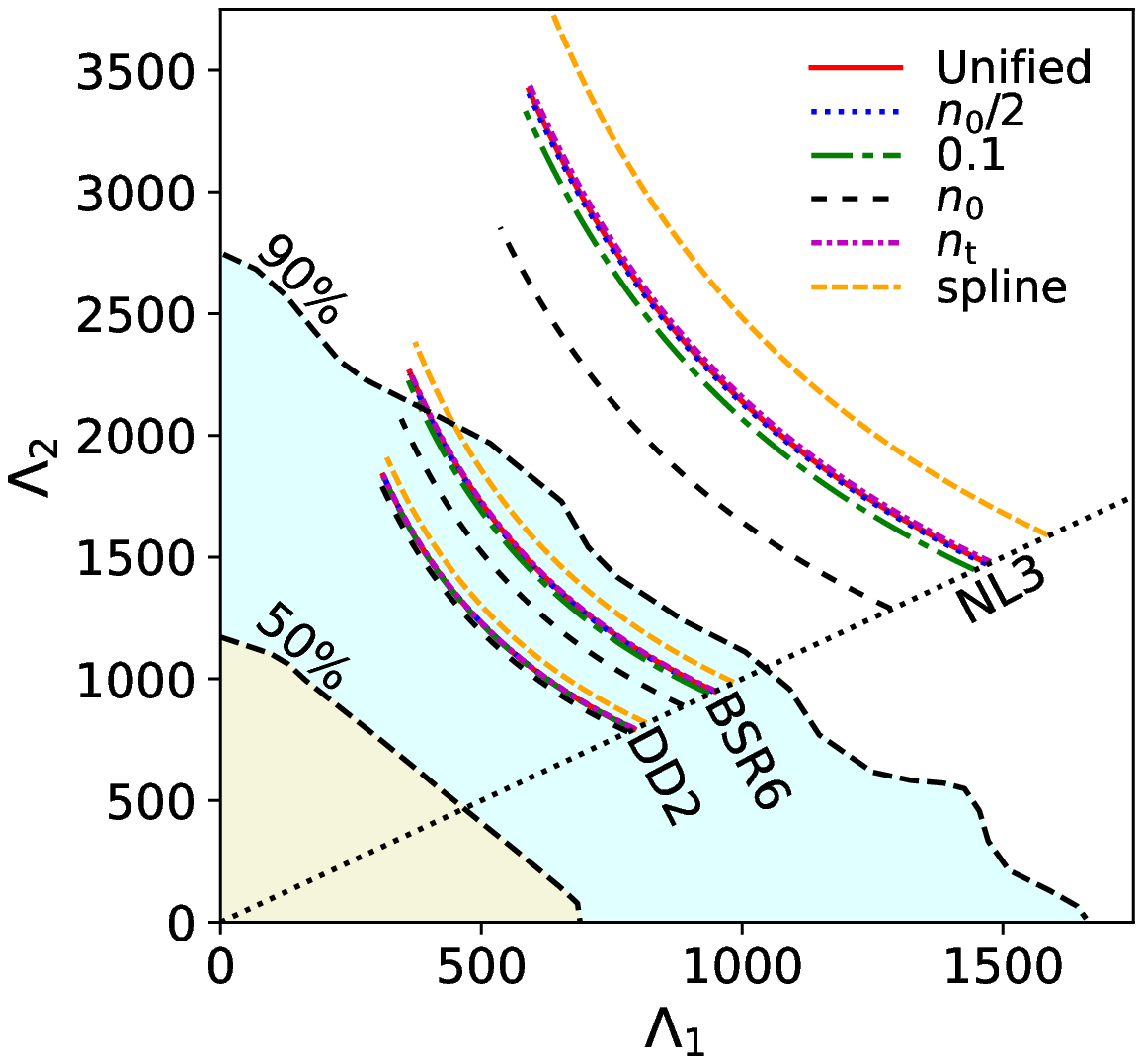}}
\caption{Influence of the matching on the tidal deformability  of the two NS of the  observed GW170817 event.}
\label{Fig:L_lambda}
\end{subfigure}
\hfill
\begin{subfigure}[b]{0.50\hsize}
\resizebox{\hsize}{!}{\includegraphics{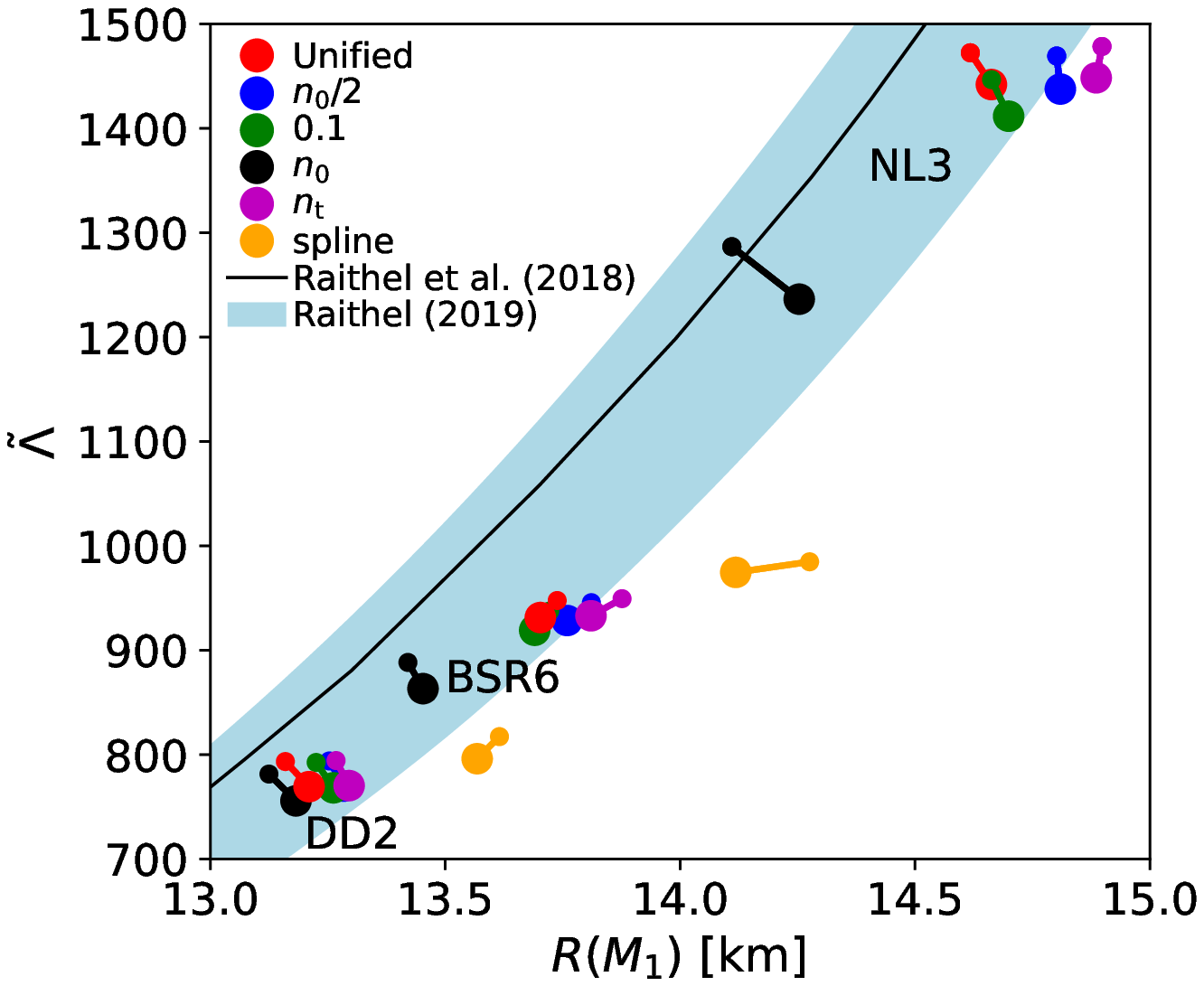}}
\caption{Influence of the matching on the relation between the effective tidal deformability and the radius of the most massive NS in GW170817. See text for details.}
\label{Fig:L_R}
\end{subfigure}
\caption{}
\end{figure*}

\subsubsection{Relations between $\Lambda$ and $I$}

We study two fits obtained in Ref.~\cite{Maselli13} and \cite{Yagi17} between $\Lambda$ and $I$ in the form: 
\begin{equation}
\ln \bar I_{\rm fit}=\sum_{k=0}^{k=4}a_k \ln(\Lambda)^k
\end{equation}
where $\bar I=c^4I/(G^2M^3)$.

The first fit from Maselli et al. \cite{Maselli13} with an indicated relative difference of less than $5\%$ yields: $a_0=1.95$, $a_1=-0.373$, $a_2=0.155$, $a_3=-0.0175$, and $a_4=0.000775$. The second one from Yagi and Yunes \cite{Yagi17} with a reported relative difference between the fit and the exact calculations of at most 1\% is given by: $a_0=1.496$, $a_1=0.05951$, $a_2=0.02238$, $a_3=-6.953\times 10^{-4}$, $a_4=8.345\times 10^{-6}$.

As for the previous fits, the relations between $M$ and $\Lambda$ together with the relative difference between the value of $I$ obtained  from the fits and from the exact calculations: $|I-I_{\rm fit}|/I_{\rm fit}$ as a function of $\lambda$ are shown in Fig.~\ref{Fig:ICLove} for the stiffest (NL3) and softest EOS (DD2) considered in this work. The value of the largest relative difference error and the NS mass at which it is reached are presented in Table \ref{tab:fit} for each EOS, matching and fit.

The relative differences between the fits and the exact calculations are similar for the three matched EOS and the unified one. The Yagi and Yunes fit gives rise to a maximum relative difference of at most $0.4\%$, well within the reported precision of the fit. However for the Maselli et al. fit, the relative difference reaches up to $7\%$ for the stiffest EOS and up to $4.5\%$ for the softer ones. These values are similar to the reported precision of each fit. The unified EOS and the three matched ones give similar values of the difference between the fits and the exact calculations. Overall the Yagi and Yunes fit performs much better than the Maselli et al. fit.

We now turn to the study of the influence of the core-crust matching on the tidal deformability in view of the recent constraints obtained from the detections of GW associated with the GW170817 event.

\begin{figure*}
\begin{subfigure}[b]{0.49\hsize}
    \resizebox{\hsize}{!}{\includegraphics{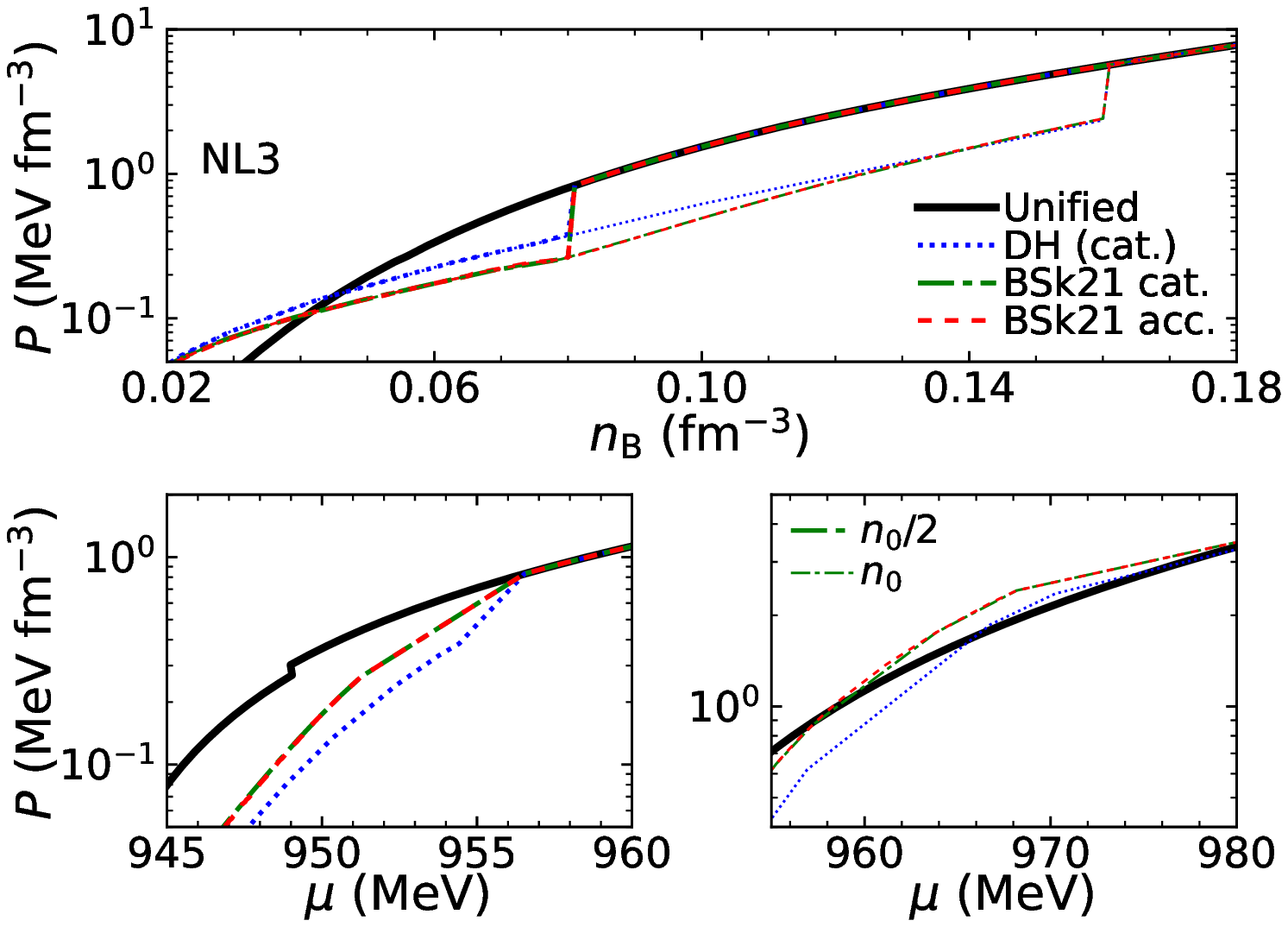}}
\caption{Matchings between NL3 core EOS and various crusts.}
\end{subfigure}
\begin{subfigure}[b]{0.49\hsize}
\resizebox{\hsize}{!}{\includegraphics{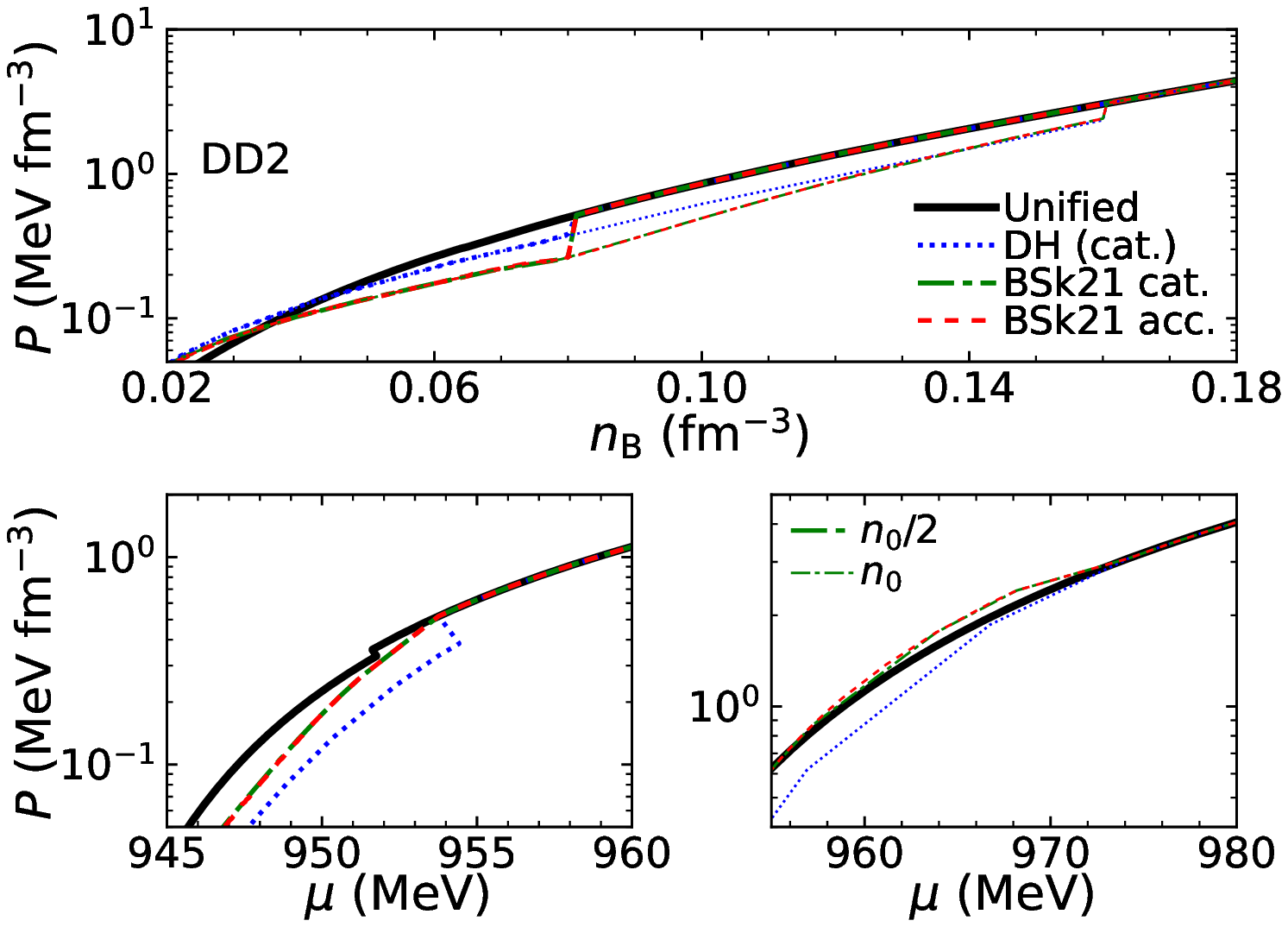}} 
\caption{Matchings between DD2 core EOS and various crusts.}
\end{subfigure}
\caption{Relations between the pressure $P$ and the baryonic density $n_B$ and the chemical potential $\mu$. Four crust EOS are employed: the EOS consistent with the crust (unified EOS), the DH and BSk21 EOSs for catalyzed matter, and the EOS for a fully accreted crust calculated for the BSk21 model. Core and crust EOSs are connected at $n_0$ (thin lines) and $n_0/2$ (thick lines).} 
\label{Fig:EOS_crust}
\end{figure*}

\subsubsection{Relations between $C$ and $I$}
Finally we study how four fits between the compactness $C$ and the moment of inertia $I$ are affected by the gluing between the core and the crust EOS.

First we consider two fits written as:
\begin{equation}
\bar I_{\rm fit}=\sum_{k=1}^4 a_k C^{-k}.
\end{equation}
The first of such fits considered here, the Yagi and Yunes fit, is derived in Ref.~\cite{Yagi17} for $ \approx 25$ NS EOSs with a reported error of at most $9\%$. It yields $a_1=1.317$, $a_2=-0.05043$, $a_3=0.04806$, and $a_4=-0.002692$. Another fit was obtained by Breu and Rezzolla \cite{Breu16} for slow-rotating models using 28 EOSs all consistent with the existence of $2\,M_\odot$ NSs with $a_1=0.8134$, $a_2=0.2101$, $a_3=0.003175$ and $a_4=-0.0002717$ with a largest deviation between the exact calculations and the fits of 3\%.

Additional fits, this time for $\tilde{I}=I/(MR^2)=\bar{I}C^2$, are also reported in the literature. For example, Zhao and Lattimer \cite{Zhao18} obtained:
\begin{equation}
\tilde{I}\simeq0.01+1.2C^{1/2}-0.1839C-3.735C^{3/2}+5.278C^2
\end{equation}
while Breu and Rezzolla \cite{Breu16} derived the following: 
\begin{equation}
    \tilde{I}=0.244+0.638C+3.202C^4,
\end{equation}
with a deviation of at most 6\%. 

The relations between $M$ and $C$ together with the relative difference between the value of $I$ obtained from fits and the exact calculations are shown in Fig.~\ref{Fig:IC} of the Appendix for the unified DD2 and NL3 EOS and three matchings of these at different densities. As before, Table \ref{tab:fit} gives for each EOS, matching, and fit the largest relative error and the corresponding value of the NS mass for which it is attained.

Interestingly the four fits, whether for $\tilde{I}$ or $\bar{I}$, have a similar precision of $5\%-6\%$ as shown in Table \ref{tab:fit}, with very little dependence on the core-crust transition density. The two fits obtained by Breu and Rezzolla appear to be more accurate (i.e., give a smaller relative error) than the two other ones
for low-mass stars $M\lesssim1.2-1.3\,M_\odot$. However these two latter fits perform better for a wider range of masses.\\

All in all similar uncertainties are obtained whether one employs the unified EOS or one matched with the crust at $n_0/2$, $n_0$ or $0.1$\,fm$^{-3}$. The Yagi and Yunes fits between $\Lambda$ and $C$ and $\Lambda$ and $I$ are not affected by the treatment of the core-crust transition, within their reported precision. The precision of the fits of Maselli et al., however, appears to be overestimated, in particular the precision of the $\Lambda-C$ fit, and turns out to be strongly affected by the matching. Finally, all fits between $C$ and $I$ are not affected by the core-crust gluing for a precision of $ \approx 6\%$.

\subsection{Consequences for GW170817 and future GW observations}

\begin{figure*}
\begin{subfigure}[b]{0.49\hsize}
\resizebox{\hsize}{!}{\includegraphics{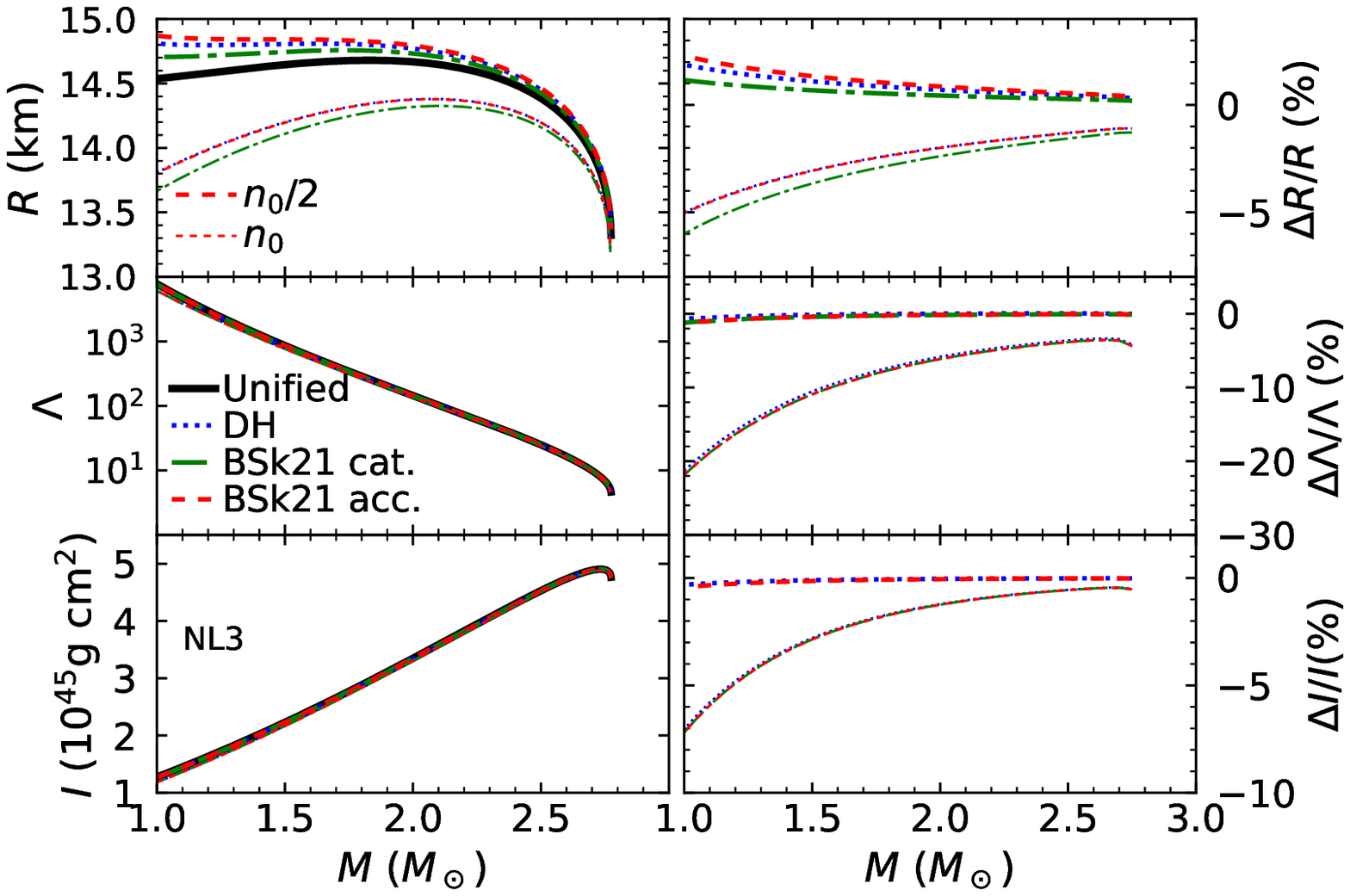}}
\caption{Macroscopic parameters for NL3 core EOS glued to various crusts.}
\end{subfigure}
\begin{subfigure}[b]{0.49\hsize}
\resizebox{\hsize}{!}{\includegraphics{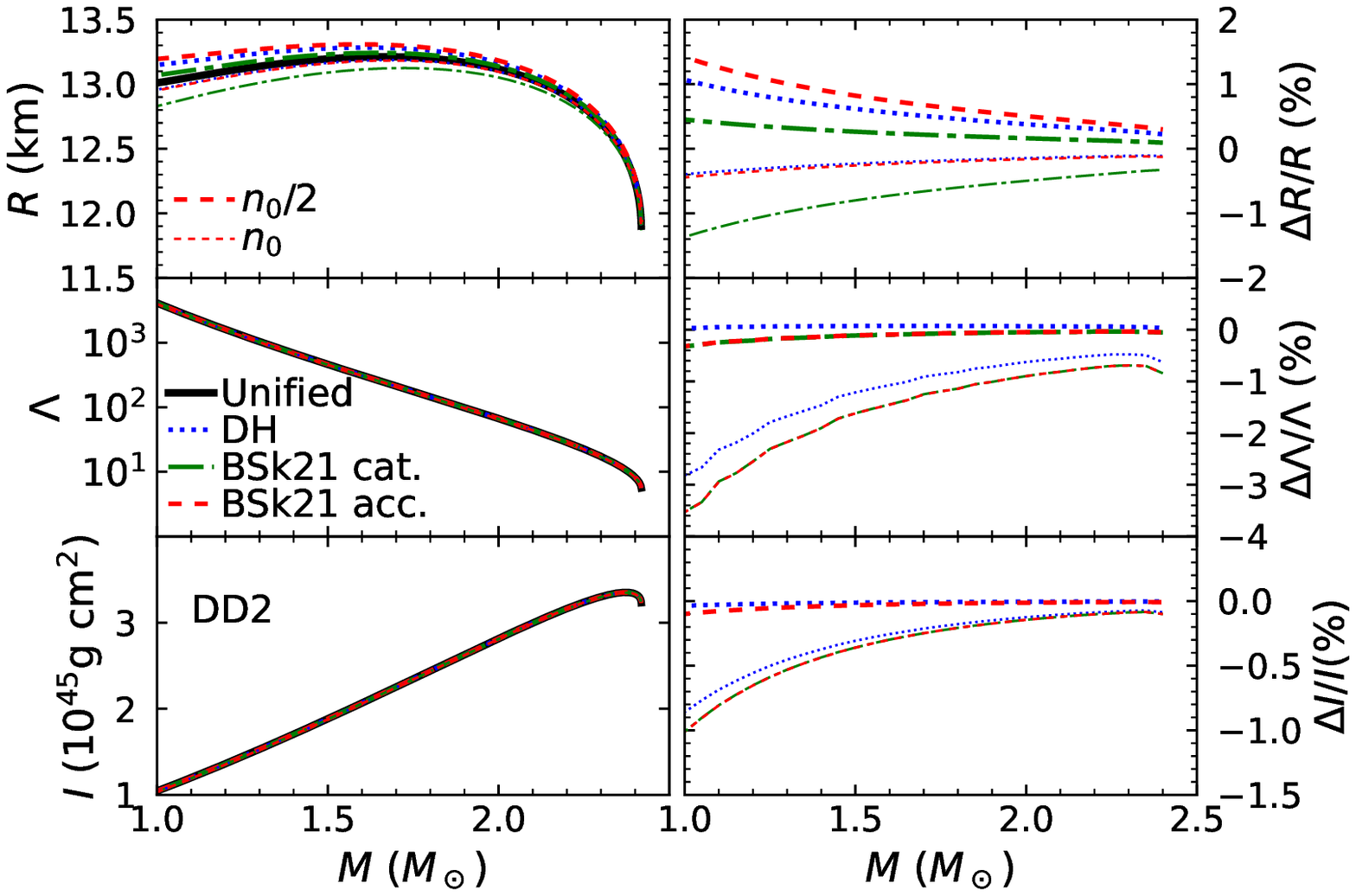}}
\caption{Macroscopic parameters for DD2  core EOS glued to various crusts.}
\end{subfigure}
\caption{$M$, $R$, $I$ and $\Lambda$ for the various EOS matchings (left) and the relative differences with respect to the unified EOS (right). Thin lines correspond to a matching between the core and the crust at $n_0$ and  the thicker ones at $n_0/2$.} 
\label{Fig:Delta_crust}
\end{figure*}

First for each core EOS and for the different matchings with the crust EOS, we calculate the tidal deformability of the two NSs while varying the mass $M_1$ of the first (heaviest) NS in the range obtained for  GW170817:  $1.365 < M_1 < 1.60\,M_\odot$. Then the mass $M_2$ of the  (lightest)  NS is determined by fixing the chirp mass $\mathcal{M}=(M_1M_2)^{3/5}(M_1 +M_2)^{-1/5}$ at its measured value: $1.188\,M_\odot$.  Results are shown in Fig.~\ref{Fig:L_lambda} together with, for reference, the 90\% and 50\% confidence limits obtained from the GW170817 for the low-spin priors as is consistent with the binary systems observed in our galaxy \cite{GW170817}. 

For the three core EOS, the $\Lambda_1-\Lambda_2$ relations obtained for   the matchings of the DH crust at $n_0/2$, $n_{\rm t}$ and $0.1$~fm$^{-3}$ and for the unified EOSs are almost indistinguishable. However one can clearly see that the curves obtained for the matching at $n_0$ and for the spline become substantially different as the stiffness of the core EOS increases. In the extreme case of the stiffest core EOS that we employ, NL3, the difference in the $\Lambda_1-\Lambda_2$ relations between the matching at $n_0$ and, say, $n_0/2$ is actually of the same order as the difference between the two core EOS, here DD2 and BSR6. As a consequence ruling out an EOS based on constraints obtained on the tidal deformability from GW observations can be impeded. Thus a careful treatment of the matching is necessary in order to rule out or not an EOS and connecting the core and crust EOS at $n_0/2$ or $0.1$ fm$^{-3}$ is recommended.

In Fig.~\ref{Fig:L_R} we explore the consequence of using a nonunified EOS on the relation between the effective tidal deformability $\widetilde\Lambda$:
\begin{equation}
\widetilde\Lambda \equiv \frac{16}{13} \frac{ (M_1 +  12 M_2) M_1^{4} \Lambda_1 + (M_2 + 12 M_1) M_2^{4} \Lambda_2}{(M_1+M_2)^5}
\end{equation}
and the radius $R(M_1)$ of the most massive NS in the binary at the origin of GW170817 as obtained in Refs.~\cite{Raithel18,Raithel_2019}. This relation is derived as being a consequence of the fact that for a NS merger, the effective tidal deformability hardly depends on the mass of the component stars for a fixed chirp mass. It then potentially allows us to use $\widetilde\Lambda$ to directly probe the NS radius. For a given EOS and each matching we compute the relation between $\widetilde\Lambda$ and $R(M_1)$ for $M_1=1.36\,M_\odot$ (indicated by the smallest dot in the figure) and $M_1=1.6\,M_\odot$ (larger dot) for a chirp mass of $1.188\,M_\odot$. We also plot as a solid black line the relation obtained in Ref.~\cite{Raithel18} for a sample of six EOSs in the form of polytropic fits. The influence of the core-crust matching is non-negligible for our three EOSs and increases strongly with the stiffness of the EOS. Hence the use of a consistent EOS appears to be required when assessing the dependence of $\widetilde\Lambda$ on the $R(M_1)$. For comparison we finally add the contour in blue corresponding to the approximate relationship obtained in Ref~\cite{Raithel_2019}. The fits between $\widetilde\Lambda$ and $R(M_1)$ obtained in Refs.~\cite{Raithel18,Raithel_2019} appear to strongly depend on the EOS matching and to be only marginally consistent with the results obtained when a unified EOS is employed.  A more-in-depth analysis is beyond the scope of this paper.

\subsection{Matching in $P-\rho$}

We also studied the influence of the matching between the core and the crust EOS but this time by gluing the two EOS not at a given baryon number density but at given mass-energy density $\rho$. For the three core EOSs we connected the core to the DH crust EOS at three densities: $\rho_0$, $\rho_0/2$, $\rho(n_{\rm B}=0.1$~fm$^{-3})$. 
For simplicity and to keep the paper short we do not include the results here. 

We obtained conclusions similar to those we got for the matching in term of $n_{\rm B}$ regarding the quality of each universal relation. Also the matchings at $\rho_0/2$ and $\rho(n_{\rm B}=0.1$~fm$^{-3})$ minimize the jump in the chemical potential as a function of the pressure and reduce the discrepancies between the macrophysical properties obtained for the matched EOS with respect to the unified one.

\section{Influence of the crust equation of state}

In the previous section we studied the influence of the matching on various NS macrophysical quantities employing a single crust EOS, namely, the DH one. We conclude that the influence of the nonconsistency of the crust and core EOSs is minimized when the two EOSs are connected at $n_0/2$ or 0.1 fm$^{-3}$. 
In this section we study the influence of the crust EOS that is employed; the relation between $\Lambda$ and $R$ calculated with and without the crust EOS was studied in Ref.~\cite{tsang2019}. We consider four crust EOSs, three catalyzed (nonaccreted) ones: the DH and BSk21 EOSs together with the crust calculated consistently with the core and a fully-accreted EOS obtained for the BSk21 nuclear model. We employ the softest and stiffest core EOSs NL3 and DD2 and connect core and crust at two densities: $n_0$ and $n_0/2$. The fully accreted crust EOS used in the present paper was based on the framework formulated in Refs.~\cite{Sato1979, Haensel1990a, Haensel1990b}, where the possibility of the neutron diffusion in the inner crust was not considered. It has recently been shown that neutron superfluidity in the inner crust could  result in the softening of the accreted crust EOS, making it closer to that of catalyzed matter, see \cite{Gusakov_2020}.

\begin{table*}
\caption{Maximum relative difference in \% between the values obtained using the various fits discussed in this paper and the exact calculations for the two BSk21 EOS, for a catalyzed (cat.) crust and an accreted (acc.) one, connected to the core at $n_0$ and $n_0/2$. Results obtained for the DH crust and the unified EOS can be found in Table \ref{tab:fit}.}
\label{tab:fit_crust}
\centering
\begin{tabular}{c|cccc|cccc}
\hline 
\hline 
EOS & cat. $n_0$ & acc. $n_0$ & cat. $n_0/2$ & acc. $n_0/2$ & $n_0$ & acc. $n_0$ & cat. $n_0/2$ & acc. $n_0/2$ \\
\hline 
 & \multicolumn{4}{c|}{} & \multicolumn{4}{c}{}\\
$\Lambda-C$ fits & \multicolumn{4}{c|}{Maselli et al.} & \multicolumn{4}{c}{Yagi and Yunes}\\
\hline 
NL3  & 2.17 (1.00)& 3.15 (1.00)& 4.96 (1.01)& 6.01 (1.01)& \textbf{3.66 (1.00)}& 2.62 (1.00)& 1.54 (1.01)& 2.16 (1.42)\\
BSR6 & 3.48 (1.01)& 4.39 (1.01)& 5.85 (1.01)& \textbf{6.80 (1.01)}& 1.22 (1.01)& 0.98 (1.33)& 2.31 (1.31)& 3.02 (1.29)\\
DD2  & 3.74 (1.01)& 4.62 (1.01)& 4.87 (1.01)& 5.76 (1.01)& 1.43 (2.20)& 1.16 (2.22)& 1.31 (1.25)&
 2.03 (1.20)\\ 
\hline 
 & \multicolumn{4}{c|}{} & \multicolumn{4}{c}{}\\

 $\bar I-\Lambda$ fits& \multicolumn{4}{c|}{Maselli et al.} & \multicolumn{4}{c}{Yagi and Yunes}\\
\hline 
NL3  & 5.06 (1.00)& 5.06 (1.00)& \textbf{7.24 (1.01)}&\textbf{ 7.24 (1.01)}& \textbf{0.41 (2.59)}& 0.40 (2.60)& 0.27 (2.65)&
 0.27 (2.65)\\
BSR6 & 4.59 (2.26)& 4.59 (2.26)& 4.49 (2.26)& 4.49 (2.26)& 0.26 (2.28)& 0.26 (2.29)& 0.17 (2.33)&
 0.17 (2.33)\\
DD2  & 4.58 (2.24)& 4.58 (2.23)& 4.54 (2.24)& 4.54 (2.24)& 0.26 (2.27)& 0.26 (2.27)& 0.22 (2.30)&
 0.22 (2.30)\\
\hline 
 & \multicolumn{4}{c|}{} & \multicolumn{4}{c}{}\\

 $\tilde{I}-C$ fits& \multicolumn{4}{c|}{Breu and Rezzolla} & \multicolumn{4}{c}{Zhao and Lattimer}\\
\hline 
NL3  & 6.39 (2.24)& 5.90 (2.37)& 4.25 (1.01)& 6.13 (1.01)& 3.62 (2.28)& 3.16 (2.41)& 3.62 (1.01)&
 5.39 (1.01)\\
BSR6 & 4.16 (2.05)& 3.66 (2.14)& 5.01 (1.01)& \textbf{6.70 (1.01)}& 1.44 (2.08)& 2.36 (1.01)& 5.08 (1.01)&
 \textbf{6.68 (1.01)}\\
DD2  & 4.60 (2.04) & 4.11 (2.12)& 3.78 (2.16)& 4.44 (1.01)& 1.88 (2.07)& 2.89 (1.01)& 3.41 (1.01)&
 4.93 (1.01)\\

\hline 
 & \multicolumn{4}{c|}{} & \multicolumn{4}{c}{}\\

 $\bar{I}-C$ fits& \multicolumn{4}{c|}{Breu and Rezzolla} & \multicolumn{4}{c}{Yagi and Yunes}\\
\hline 
NL3  & \textbf{5.34 (2.76)}& 5.07 (2.76)& 3.69 (2.77)& 3.42 (2.77)& 3.90 (1.00)& 3.03 (2.13)& 3.25 (2.51)&
 3.45 (2.51)\\
BSR6 & 2.68 (2.39)& 2.36 (2.40)& 2.19 (1.01)& 3.82 (1.01)& 1.46 (1.87)& 1.07 (1.12)& 3.56 (1.16)&
 \textbf{4.89 (1.13)}\\
DD2  & 3.21 (2.39) & 2.90 (2.40)& 2.65 (2.41)& 2.35 (2.41)& 1.84 (1.89)& 1.99 (2.32)& 2.17 (2.32)&
 3.48 (1.04)\\
\hline 
\hline 
\end{tabular} 
\end{table*}

\subsection{Neutron-star macrophysical properties}
In Fig.~\ref{Fig:EOS_crust} we show the pressure as a function of the density and chemical potential when the different crusts are glued to the core at $n_0$ and $n_0/2$, for the two core EOS. We note that the two BSk21 EOS for a nonaccreted and fully accreted crust give almost the same relation between the pressure and the density and chemical potential for the range of values presented in Fig.~\ref{Fig:EOS_crust}. This is however not the case at lower densities.  Indeed the various reactions (electron captures followed by pycno-nuclear reactions as the density increases) that take place in an accreted crust are each accompanied by a jump in the chemical potential as a function of pressure. It has been shown in Ref.~\cite{Fantina18} that although the composition of accreted and catalyzed crusts are different, the main parameters connected with pressure and the equation of state (proton fraction and fraction of nucleons in nuclei) are converging at pressure $ \approx 0.01~{\rm MeV\,fm}^{-3}$, therefore below values presented in Fig.~\ref{Fig:EOS_crust}. 
As in the previous section at the core-crust transition, the jumps in the chemical potential as a function of the pressure due to the use of a nonunified EOS are much smaller when the gluing is performed at $n_0/2$ than at $n_0$ whichever core EOS is used. Comparing the BSk21 and DH EOS, we observe that for the latter the jumps in $\mu$ are larger than for the former for a gluing at $n_0/2$ while they are roughly of the same order at $n_0$.

In Fig. \ref{Fig:Delta_crust} we show, for the different crusts glued to the NL3 and DD2 core at $n_0/2$ and $n_0$, the relations between the mass, the radius, the tidal deformability and the moment of inertia, and the relative difference with respect to the unified EOS. In line with the previous section, the matching of the core and the crust EOS at $n_0$ gives much larger relative differences in $R$, $\Lambda$ and $I$ than the one at $n_0/2$. The calculations of the three macrophysical properties appear to hardly depend on the choice of the crust EOS. The main contribution to the uncertainty actually originates from the choice of the matching density and very little from the crust EOS itself. For the stiffest core EOS NL3 glued at $n_0/2$, the difference is at most of $ \approx 2.5\%$ for the radius, $ \approx 1\%$ in the deformability and $0.5\%$ in the moment of inertia while it can reach $5\%$, $20\%$, and $7$\% for a matching at $n_0$. In fact, the crust EOS is the part of a NS that is the best constrained from nuclear experiments and measurements of properties of nuclei. The two models DH and BSk21 have been calibrated to reproduce the properties of a large number of nuclei. Consequently these two models give very similar results for the case of a catalyzed crust. The case of an accreted crust is different because the density jumps it exhibits directly relate to an increase of the radius (see discussion in Ref.~\cite{Zdunik17}). This is why the accreted crust gives larger relative differences compared with the unified EOS. However, in the end the choice of the core-crust transition influences results in larger differences with respect to the unified EOS than the use of a catalyzed or an accreted crust.

\begin{figure*}
\begin{subfigure}[b]{0.46\hsize}
\resizebox{\hsize}{!}{\includegraphics{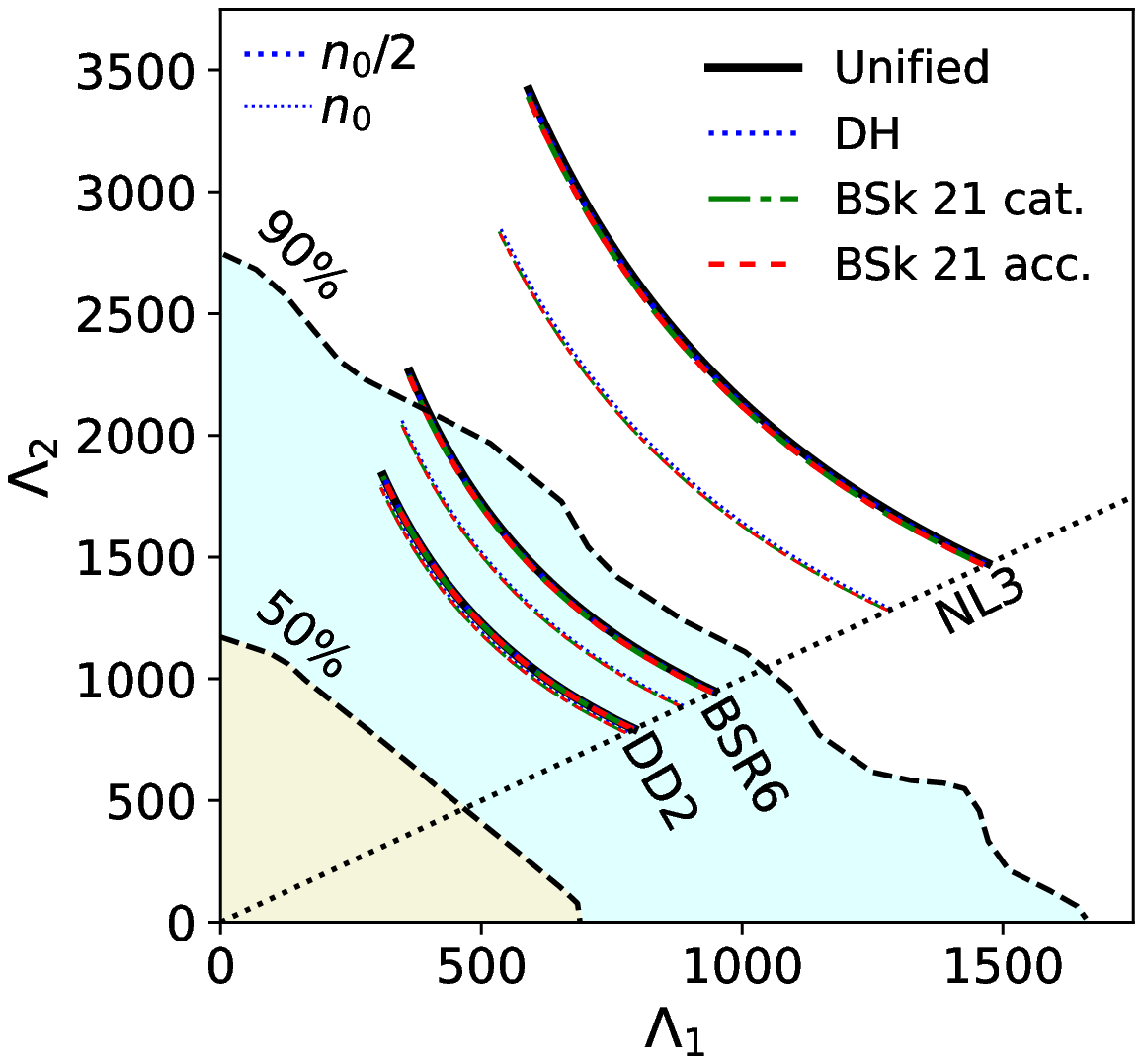}}
\caption{Influence of the crust EOS on the tidal deformability  of the two NS of the  observed GW170817 event.}
\label{Fig:L_lambda_crust}
\end{subfigure}
\hfill
\begin{subfigure}[b]{0.53\hsize}
\resizebox{\hsize}{!}{\includegraphics{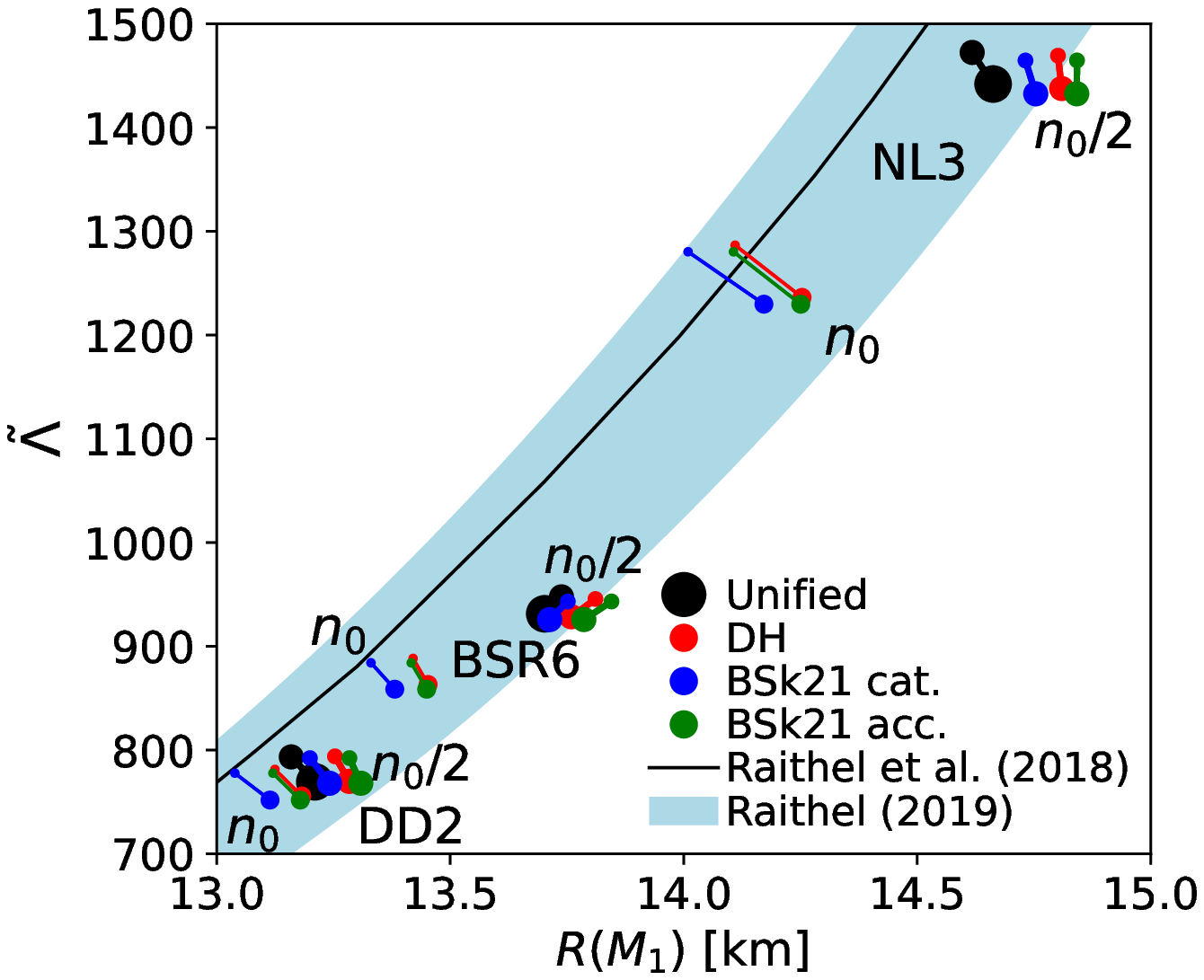}} 
\caption{Influence of the crust EOS on the relation between effective tidal deformability and the radius of the most massive NS in GW170817.}
\label{Fig:L_R_crust}
\end{subfigure}
\caption{}
\end{figure*}

\subsection{Universal relations}

We now turn to universal relations. In Figs.~\ref{Fig:ICLove_crust} and \ref{Fig:IC_crust} of the Appendix and in Table \ref{tab:fit_crust}, we compare the precision of the fits $C-\Lambda$, $I-\Lambda$ and $C-I$ relations obtained for the BSk21 catalyzed and accreted crusts models connected to the core at $n_0/2$ and $n_0$ and for the unified EOS. Results for the DH and the unified crusts can be found in Table \ref{tab:fit}.

Similarly to the conclusions drawn in the previous section, the fit between  $C$ and $\Lambda$ obtained by Yagi and Yunes \cite{Yagi17} has a precision of $ \approx 3\%-4\%$ well below the reported precision while the one presented by Maselli et al. \cite{Maselli13} gives rise to uncertainties as large as $7\%$. The latter fit is better for masses $ \approx2\,M_\odot$ and above. Overall the precision is better for the EOS matched at $n_0/2$ than at $n_0$ whichever crust model is used. The matching with the BSk21 accreted crust gives rise to the largest discrepancies between the fit and the exact calculations.

The fits obtained between $I$ and $\Lambda$ by Yagi and Yunes are overall about an order of magnitude better than those from Maselli et al., the relative deviations reaching at most $1\%$ for the former and $7\%$ for the latter. The matchings at $n_0/2$ of the BSk21 crusts are the less accurate ones.

As far as the relations between the moment of inertia $I$ and the compactness $C$ are concerned, those obtained between $\tilde{I}$ and $C$ and between $\bar{I}$ and $C$ give similar precisions, of the order of $7\%$. The Yagi and Yunes fit between $\bar{I}$ and $C$ are more accurate than that obtained by Breu and Rezzolla \cite{Breu16}, while the fit between $C$ and $\tilde{I}$ obtained in the latter reference is more precise than that presented by Zhao and Lattimer \cite{Zhao18}, except for low mass stars $M\lesssim 1.3\,M_\odot$.

All in all, for a given fit, similar precisions are obtained whichever crust EOS are used. Thus the choice of the crust EOS hardly affects the universal relations between $I$, $C$, and $\Lambda$ and we confirm the conclusion of the previous section regarding the precision of each universal relation considered in this study.

\subsection{Consequences for GW170817 and future GW observations}

We explore the consequence of the crust EOS on the relation between the tidal deformability of the two NSs that merged generating GW170817 in Fig.~\ref{Fig:L_lambda_crust} and on the relation between the effective tidal deformability $\tilde{\Lambda}$ and the radius of the most massive NS in the binary at the origin of GW170817 in Fig.~\ref{Fig:L_R_crust}. We confirm the fact that the matching of the crust to the core at $n_0/2$ gives results that are closer to those with the unified EOS than the matching at $n_0$. Whichever crust model is used the relations between $\Lambda_1$ and $\Lambda_2$ are indistinguishable. The relations are, however, dependent on the core-crust transition density in line with the results of the previous section and the effect can be as large as the difference between two EOS. We note that the relation between $\widetilde\Lambda$ and $R(M_1)$ depends much less on the nature of the crust than on the location of the core-crust interface.

\section{Conclusions}

We conclude by pointing out the importance of ensuring thermodynamic consistency in the construction of the equation of state when one wants to model the macroscopic parameters of neutron stars. A jump in the chemical potential at the core-crust interface should be as small as possible when employing nonunified EOS. This can be achieved by gluing core and crust at a density in the range of $0.08-0.1$ fm$^{-3}$. This is a direct consequence of the fact that nuclear models are adjusted to reproduce the results of laboratory experiments which constrain the property of matter up to roughly half the nuclear saturation density. We show that matching the core and crust EOSs at another density can create a relative difference with respect to the unified EOS as large as $5\%$ for the radius, $20\%$ for the deformability, and $10\%$ for the moment of inertia.
As far as universal relations $\Lambda-C$ and $\Lambda-I$ are concerned, the reported precision of the fits obtained by Maselli et al. is overestimated and strongly affected by the use of a non-unified EOS while the ones derived by Yagi and Yunes give very good results. All considered universal relations between $C-I$ perform equally well. Finally the density of the core-crust matching can result in a discrepancy with respect to a unified crust in the relation $\Lambda_1-\Lambda_2$ that can be as large as those obtained for different core EOSs. It is worth noting however that this  $\Lambda_1-\Lambda_2$ is hardly sensitive to the use of a catalyzed or accreted crust showing that inferring properties of the crust from such relations is not possible. Similar trends are obtained in the relation between the effective tidal deformability and the radius of the most massive NS in the binary.

We also point out the need to use unified EOSs in order to calculate the macroscopic properties of NSs. A number of current works rely on the polytropic fits from Ref.~\cite{Read09} for its simplicity of use. However, to obtain these fits, a single crust, the DH crust, has been employed for over 30 core models and thus the fitted EOSs are not consistent. In a follow up presentation, we will revise polytropic fits, based on unified models.

\begin{acknowledgments}
We thank C. Providencia and M. Bejger for useful discussions and comments and N. Chamel for providing the nuclear parameters of the BSk21 below saturation.
The authors acknowledge the financial support of the National Science Center Poland grant 2017/26/D/ST9/00591 (MF and LS) and grant 2018/29/B/ST9/02013 (LS, JLZ and PH).
\end{acknowledgments}

\providecommand{\noopsort}[1]{}\providecommand{\singleletter}[1]{#1}%

\newpage

\appendix
\section{Appendix}
\begin{figure*}[h!]
\begin{subfigure}[b]{0.47\hsize}
\resizebox{\hsize}{!}{\includegraphics{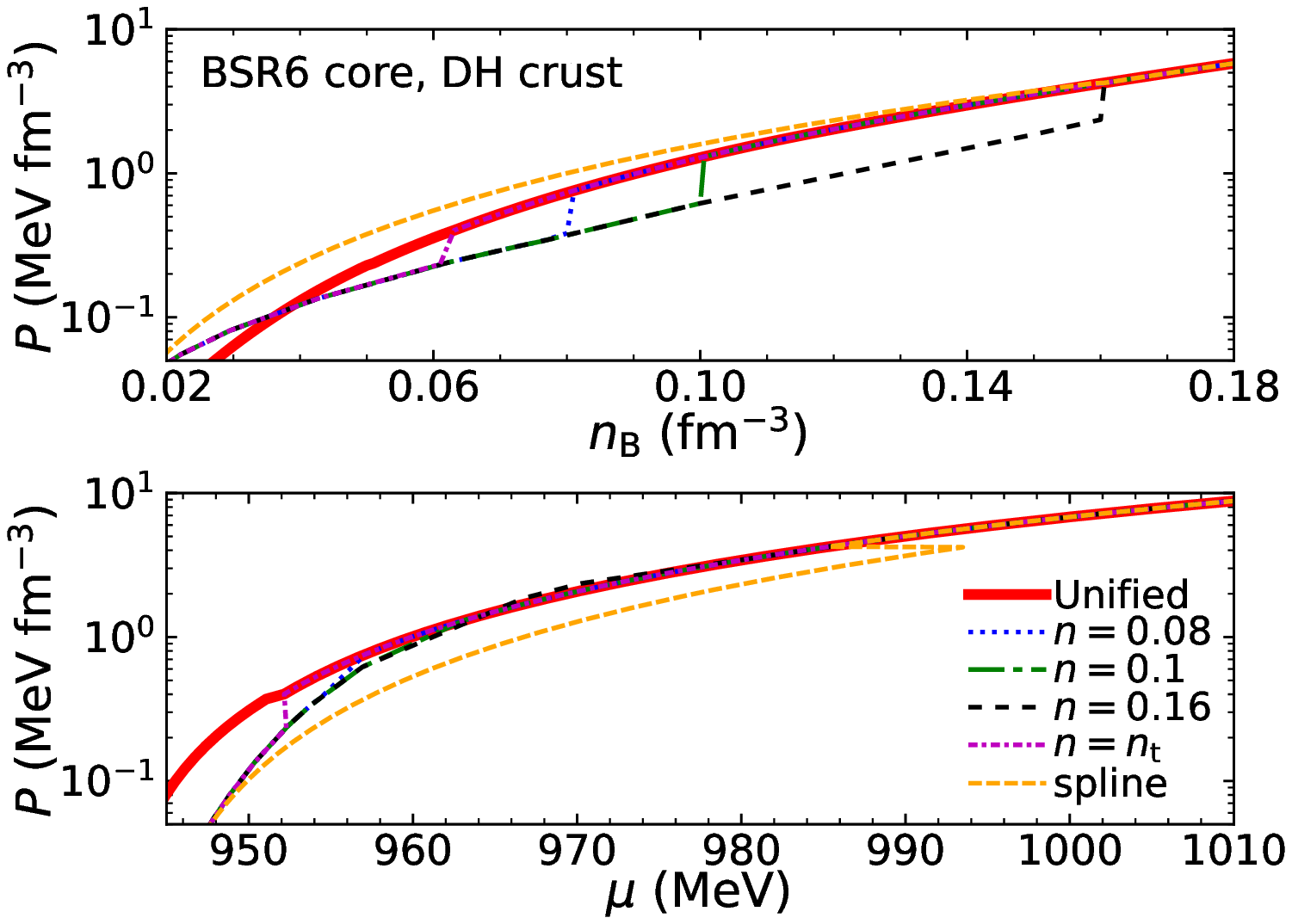}}
\caption{Pressure $P$ as a function of the baryon number density $n_B$ (upper plot) and the chemical potential $\mu$ (lower plot) for the various matched and unified EOS with the BSR6 core model.}
\label{Fig:EOS_BSR6}
\end{subfigure}
\hfill
\begin{subfigure}[b]{0.52\hsize}
\resizebox{\hsize}{!}{\includegraphics{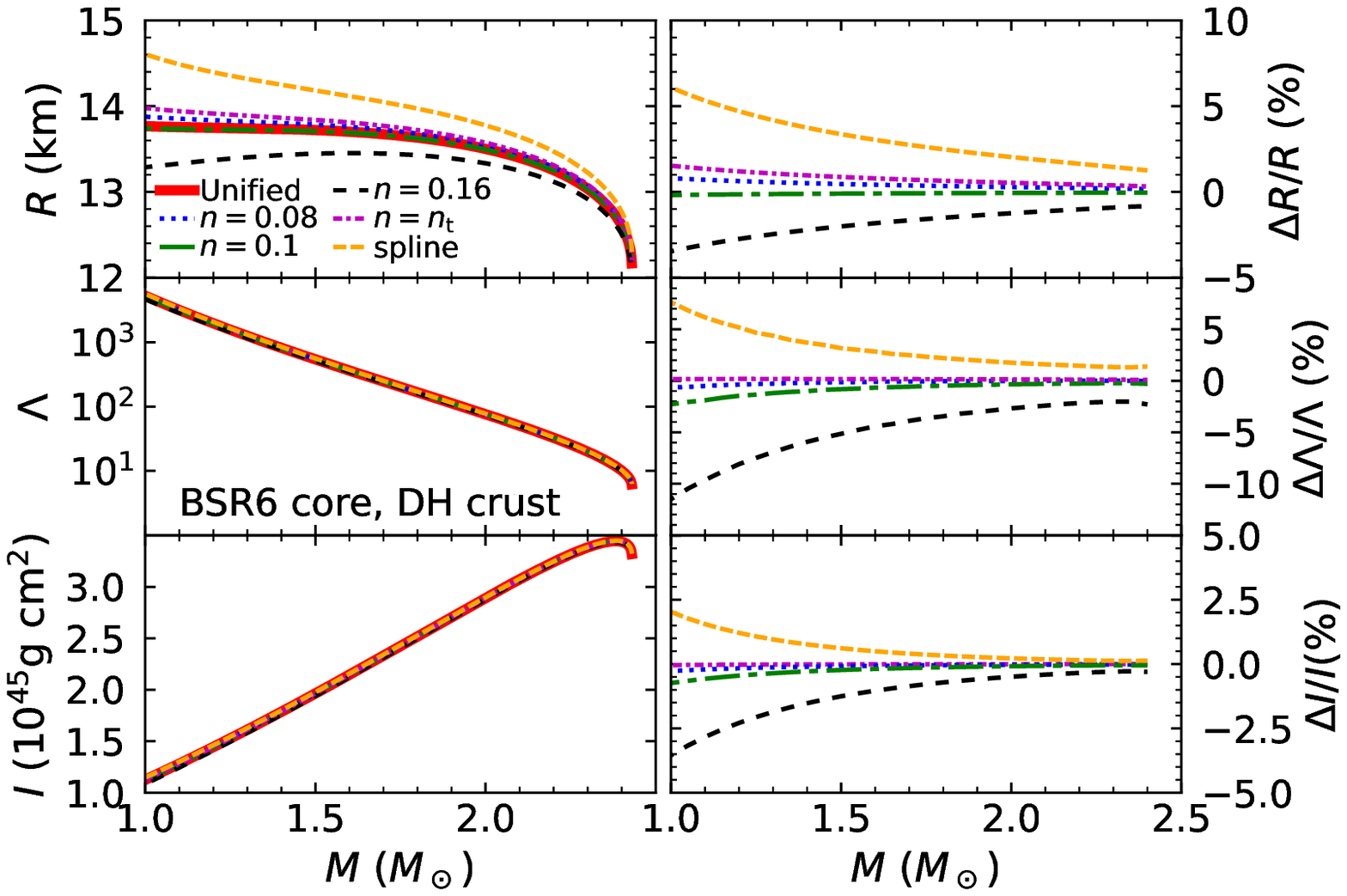}} 
\caption{$M$, $R$, $I$ and $\Lambda$ for the various matched and unified EOS (left) and the relative differences with respect to the unified EOS BSR6 (right).}
\label{Fig:Delta_BSR6}
\end{subfigure}
\caption{Matchings between the crust and the BSR6 core EOS and ensuing uncertainties on macrophysical parameters.}
\end{figure*}

\begin{figure*}[h!]
\resizebox{\hsize}{!}{\includegraphics{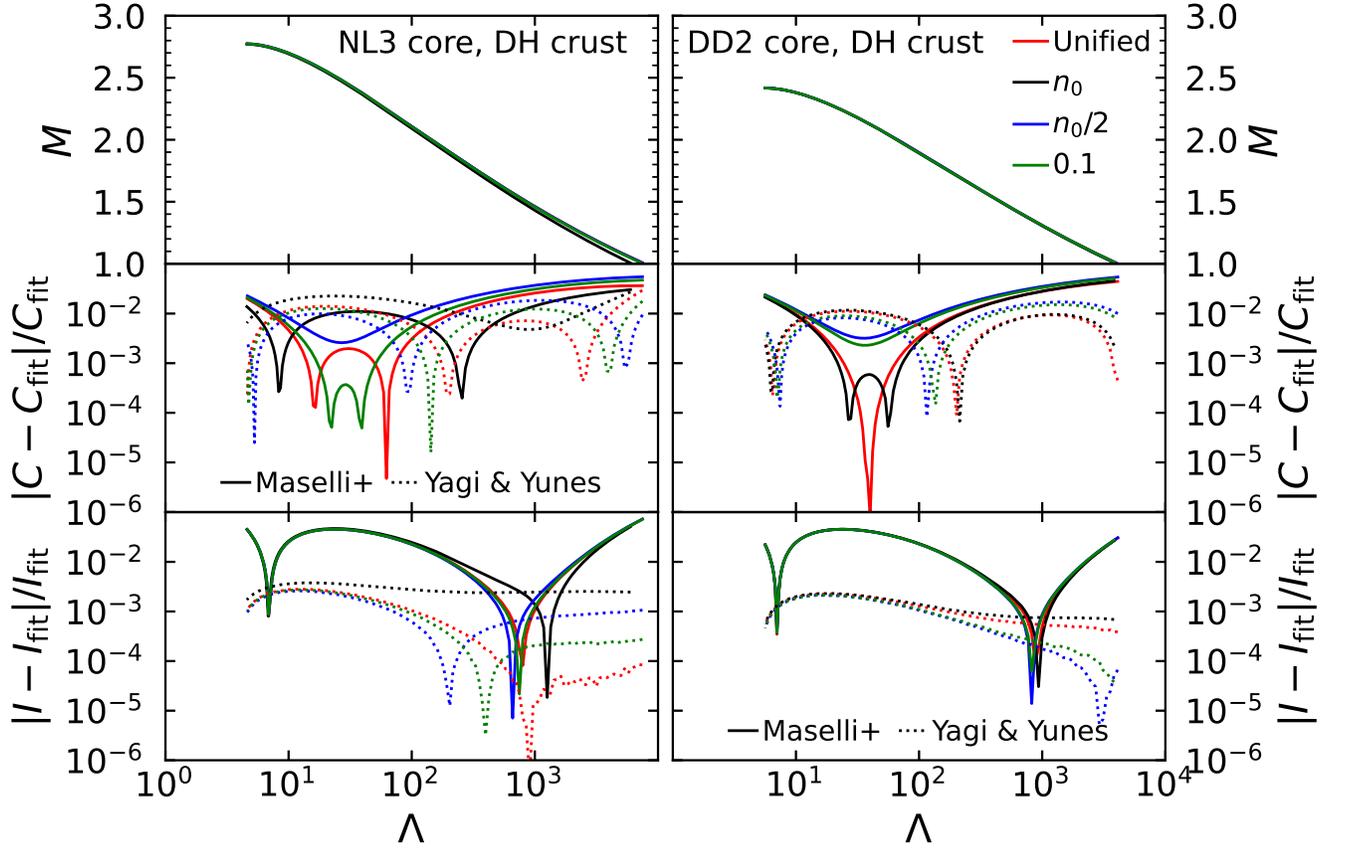}}
\caption{$C-\Lambda$ and $I-\Lambda$ fits: for the NL3 (left plots) and the DD2 (right plots) core EOS, relations between $\Lambda$ and $M$  in the top panel and relative error between the Maselli et al. (solid lines) and  the Yagi and Yunes (dotted lines) $C-\Lambda$ and $I-\Lambda$  fits with respect to exact calculations. In addition to the unified EOS, results are shown for the DH EOS connected to the core EOS at $n_0$, $n_0/2$, and $0.1$ fm$^{-3}$.}
\label{Fig:ICLove}
\end{figure*}

\begin{figure*}[h!]
\resizebox{\hsize}{!}{\includegraphics{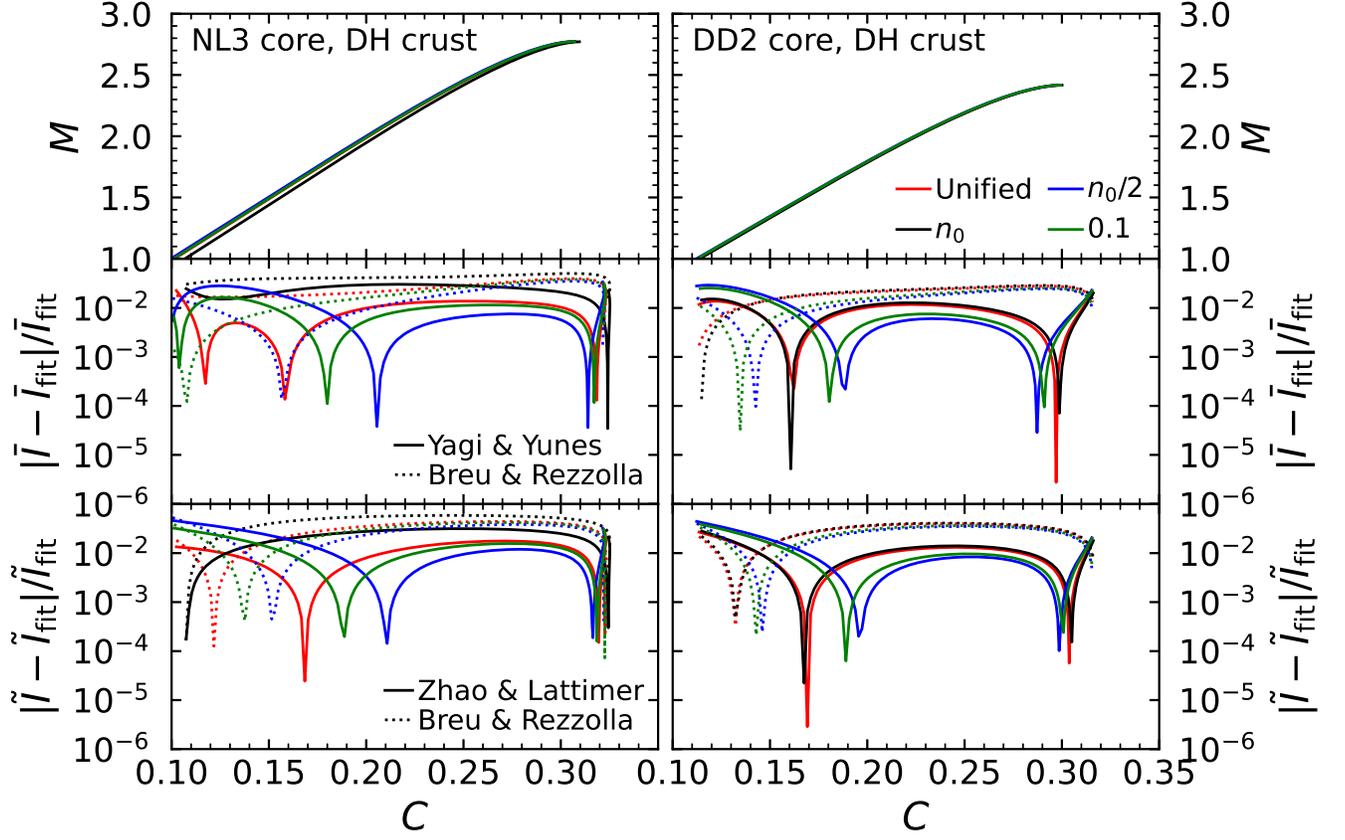}}
\caption{$C-I$ fits: for the NL3 (left plots) and the DD2 (right plots) core EOS, relations between $C$ and $M$  in the top panel and relative error for the four fits between $C$ and $I$ discussed in the text. In addition to the unified EOS, results are shown for the DH EOS connected to the core EOS at $n_0$, $n_0/2$, and $0.1$ fm$^{-3}$.}
\label{Fig:IC}
\end{figure*}

\begin{figure*}[h!]
\resizebox{\hsize}{!}{\includegraphics{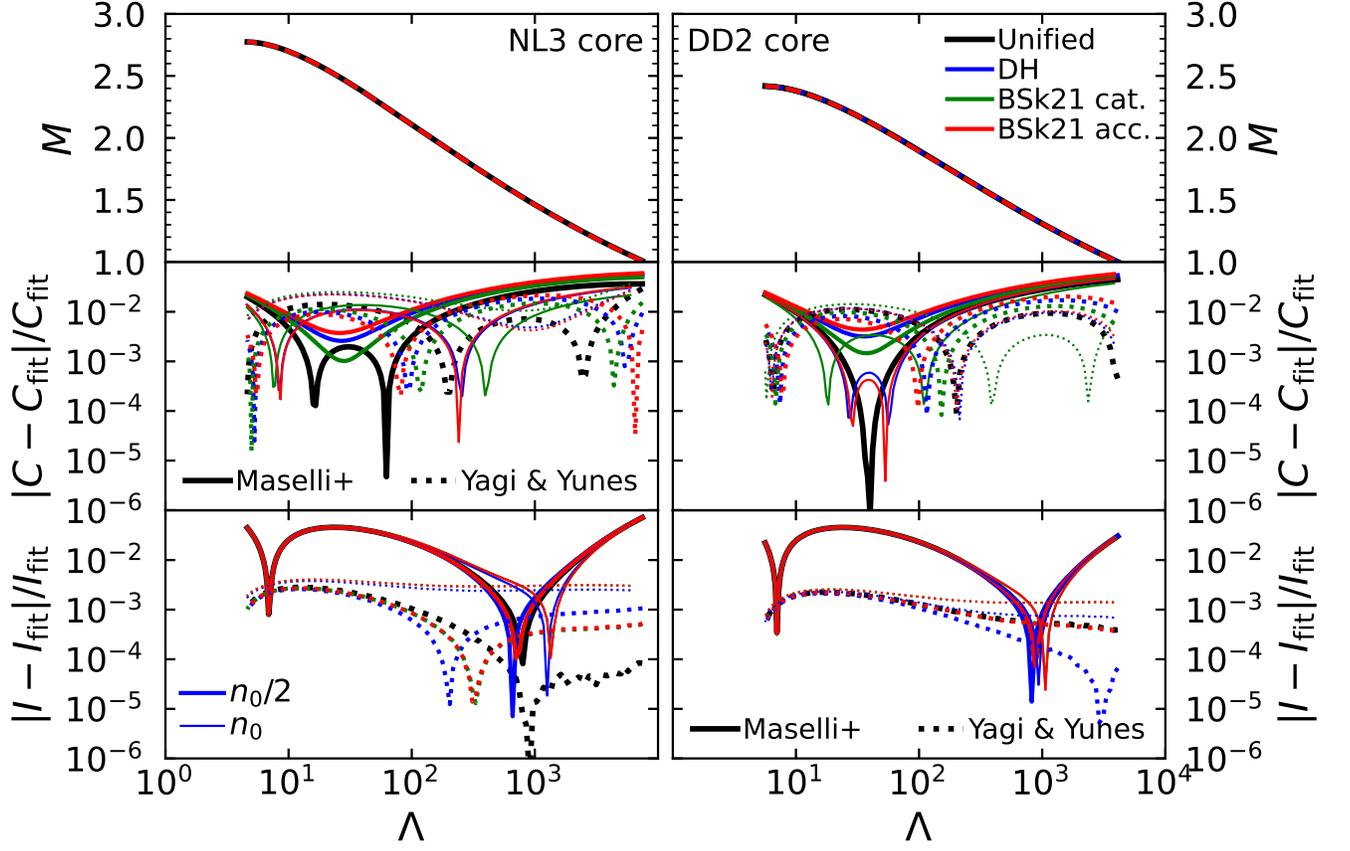}}
\caption{$C-\Lambda$ and $I-\Lambda$ fits: for the NL3 (left plots) and the DD2 (right plots) core EOS, relations between $\Lambda$ and $M$  in the top panel and relative error for the Maselli et al. (solid lines) and  the Yagi and Yunes (dotted lines) $C-\Lambda$ and $I-\Lambda$  fits. In addition to the unified EOS, results are shown for the three different crust EOS connected to the core EOS at $n_0$ and $n_0/2$.}
\label{Fig:ICLove_crust}
\end{figure*}

\begin{figure*}[h!]
\resizebox{\hsize}{!}{\includegraphics{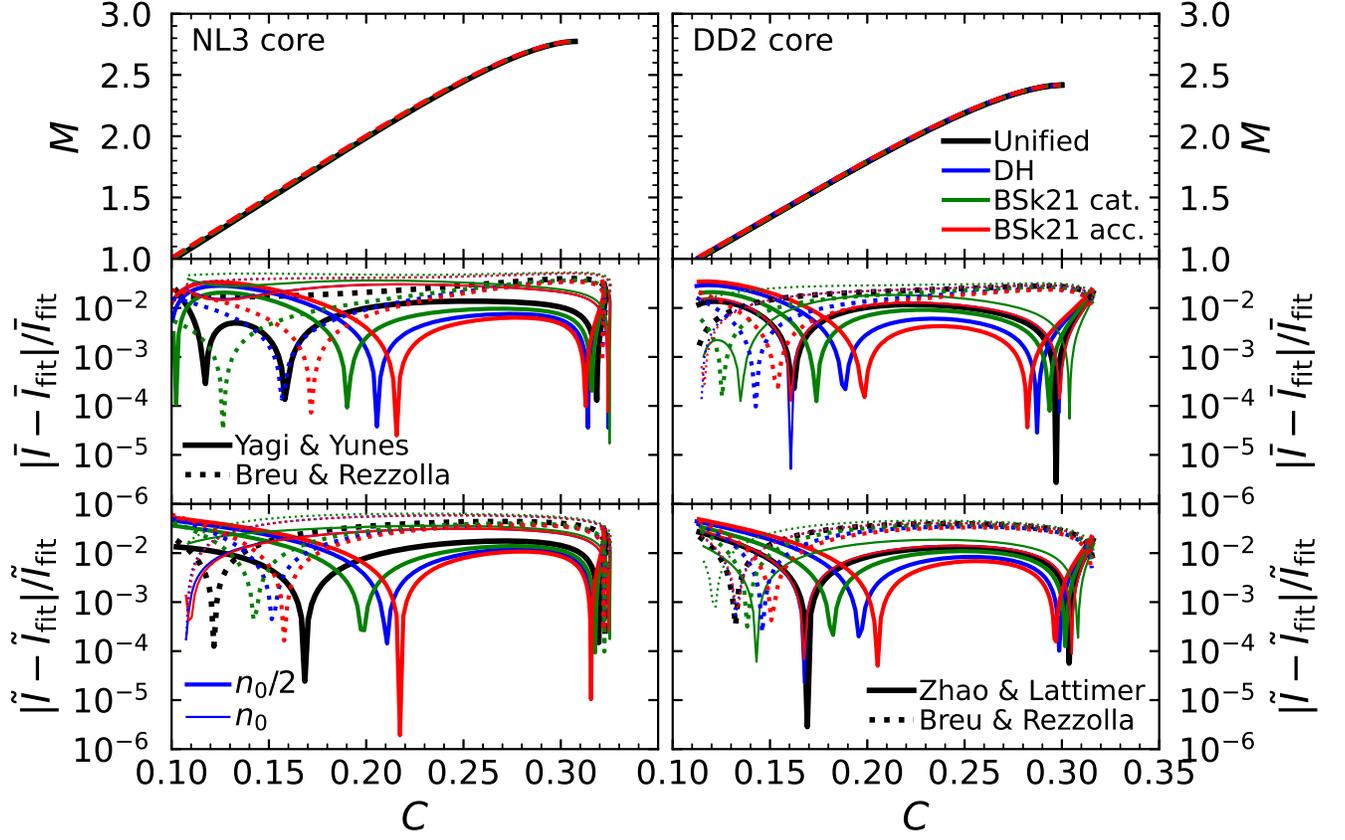}}
\caption{$C-I$ fits: for the NL3 (left plots) and the DD2 (right plots) core EOS, relations between $C$ and $M$  in the top panel and relative difference between the four fits between $C$ and $I$ discussed in the text. In addition to the unified EOS, results are shown for the three different crust EOSs connected to the core EOS at $n_0$ and $n_0/2$.}
\label{Fig:IC_crust}
\end{figure*}

We present additional figures that were used to support the study: 
\begin{enumerate}
    \item Results of matching core and crust for BSR6 EOS, which stiffness lies between that of DD2 and NL3 EOS are presented in Fig.~\ref{Fig:EOS_BSR6} (equation of state construction) and Fig.~\ref{Fig:Delta_BSR6} (uncertainties on macroscopic parameters).
    \item Uncertainty on macroscopic parameters related to various fits discussed in Sec. III C for DD2 and NL3 EOS glued to DH crust are presented in Fig.~\ref{Fig:ICLove} for universal relations $C-\Lambda$  and $I-\Lambda$ and in Fig.~\ref{Fig:IC} for universal relation $C-I$.
    \item Uncertainty on macroscopic parameters related to various fits discussed in Sec. IV B for DD2 and NL3 EOS glued to BSK21 catalyzed and accreted crusts are presented in Fig.~\ref{Fig:ICLove_crust} for universal relations $C-\Lambda$ and $I-\Lambda$  and in Fig.~\ref{Fig:IC_crust} for universal relation $C-I$. 
\end{enumerate}

\end{document}